\documentclass[]{aastex631}
\usepackage{natbib}  % to use \citet
\usepackage{graphicx}	% Including figure files
\usepackage{amsmath}	% Advanced maths commands
\usepackage{amssymb}	% Extra maths symbols

\usepackage{amsmath}
\usepackage{bm}
\usepackage{graphicx}	% Including figure files
\usepackage{grffile}
\usepackage{amsmath}	% Advanced maths commands
\usepackage{amssymb}	% Extra maths symbols

\usepackage{epsfig,amsmath,natbib}
\usepackage{color,subfigure}
\usepackage{xcolor}

\usepackage{mathrsfs,amssymb,amstext}
\usepackage{ulem}
\usepackage{url}
\usepackage{bm}
\usepackage{gensymb} % to use \degree commend
\usepackage{textcomp}

\usepackage[T1]{fontenc}

\DeclareRobustCommand{\VAN}[3]{#2}
\let\VANthebibliography\thebibliography
\def\thebibliography{\DeclareRobustCommand{\VAN}[3]{##3}\VANthebibliography}

\def\black{\color{blue}}

\def\black{\color{black}}

\usepackage{booktabs}

\begin{document}

\title{Loop I/NPS morphology predictions in the ultralong-wavelength band}

\correspondingauthor{Bin Yue}
\email{yuebin@nao.cas.cn}

\author{Yanping Cong}
\affiliation{Shanghai Astronomical Observatory, Chinese Academy of Sciences, Shanghai 200030, China}

\author{Bin Yue}
\affiliation{National Astronomical Observatories, Chinese Academy of Sciences, 20A Datun Road, Chaoyang District, Beijing 100101, China}
\affiliation{Key Laboratory of Radio Astronomy and Technology, Chinese Academy of Sciences, 20A Datun
Road, Chaoyang District, Beijing 100101, China}

\author{Yidong Xu}
\affiliation{National Astronomical Observatories, Chinese Academy of Sciences, 20A Datun Road, Chaoyang District, Beijing 100101, China}
\affiliation{Key Laboratory of Radio Astronomy and Technology, Chinese Academy of Sciences, 20A Datun
Road, Chaoyang District, Beijing 100101, China}

\author{Furen Deng}
\affiliation{National Astronomical Observatories, Chinese Academy of Sciences, 20A Datun Road, Chaoyang District, Beijing 100101, China}
\affiliation{School of Astronomy and Space Science, University of Chinese Academy of Sciences, Beijing 100049, China}

\author{Jiajun Zhang}
\affiliation{Shanghai Astronomical Observatory, Chinese Academy of Sciences, Shanghai 200030, China}

\author{Xuelei Chen}
\affiliation{National Astronomical Observatories, Chinese Academy of Sciences, 20A Datun Road, Chaoyang District, Beijing 100101, China}
\affiliation{Key Laboratory of Radio Astronomy and Technology, Chinese Academy of Sciences, 20A Datun
Road, Chaoyang District, Beijing 100101, China}
\affiliation{School of Astronomy and Space Science, University of Chinese Academy of Sciences, Beijing 100049, China}

\shorttitle{Loop I/NPS at ultra-long wavelength}
\shortauthors{Cong et al.}



\begin{abstract}

Loop I/North Polar Spur (NPS) is the giant arc structure above the Galactic plane  
{\black observed at radio wavelengths ($\lesssim 10$ GHz)}. 
There has been long-standing debate about its origin. While many people believe that it
{\black consists of}
nearby supernova remnants (SNRs), some others consider it as a giant bubble close to the Galactic Center (GC), associated with the Fermi Bubble and the eROSITA X-ray bubble.  
{\black At ultra-long wavelengths} 
 (wavelength $\gtrsim 10$ m or frequency $\lesssim 30$ MHz), particularly below $\sim 10$ MHz, the free-free absorption of the radio signal by diffuse electrons in the interstellar medium (ISM) becomes significant, resulting in different sky morphologies from those at higher frequencies. In this paper, {\black we develop emissivity models for the two Loop I/NPS origin scenarios, and predict the Loop I/NPS morphology at ultra-long {\black wavelengths} in both scenarios, taking into account the free-free absorption effect.
We find that in the SNRs model, the full Loop I/NPS will still be 
a bright arc even at  $\sim 1$ MHz. In the GC model, the arc is fully visible only above $\sim 3$ MHz. While below this frequency, it is visible only at Galactic latitudes $b\gtrsim 30\degree$; the $b\lesssim 30 \degree$ part becomes invisible due to the absorption by the ISM electrons between the GC and the Sun.} The upcoming space missions aiming at ultra-long wavelengths, such as the DSL and the FARSIDE, can potentially distinguish these two scenarios and provide decisive information about the origin of the Loop I/NPS.

\end{abstract}



\section{Introduction}

In radio bands, Loop I exhibits a conspicuous giant loop-like (arc) feature above the Galactic plane in the sky \citep{Baldwin1955MNRAS,Roger1999A&A,Dowell2017MNRAS,Eastwood2018AJ,Guzman2011A&A,Landecker1970AuJPA,Patra2015ApJ,Haslam1982A&A,Reich1986A&A,IntemaHT2017A&A}, covering an area of $\sim 70\degree \times 50\degree$ \citep{Lallement2023CRPhy}. {\black It crosses the Galactic plane} at Galactic longitude $l\sim30$ \degree and Galactic latitude $b\sim10$\degree, 
and extends toward the north Galactic pole to about $l\sim330$\degree and $b\sim80$\degree. 
There is {\black also a bright} spur near {\black the root of} Loop I \citep{HanburyBrown1960Obs}, named North Polar Spur (NPS, e.g. \citealt{Iwashita2023ApJ}). 
In this paper, we use ``Loop I/NPS'' to refer to the full loop-like structure. 
In the soft X-ray band, there is also a loop-like structure above the Galactic plane, {\black which coincides largely} with the radio Loop I/NPS \citep{Egger1995A&A}. 
The X-ray structure is quasicircular and is generally identified as a bubble, i.e. the northern eROSITA bubble. 
Below the Galactic plane, there is the southern eROSITA bubble that is roughly symmetric to its sibling  
\citep{Predehl2020Natur}.
However, in the radio bands, no corresponding structure was found in the total intensity map. 

Besides the Loop I/NPS, other loop-like structures have also been found and designated as Loop II (Cetus arc) \citep{Large1962MNRAS}, Loop III \citep{Quigley1965StructureOT}, and Loop IV \citep{Large1966MNRAS,Berkhuijsen1971A&A}, respectively.
The Loop I structure, and many other smaller and dimmer loop-like structures, are highly polarized \citep{Planck2016XXV}. The polarized structure of Loop I seems to extend far below the Galactic plane \citep{Vidal2015MNRAS,Panopoulou2021ApJ}.

The debate on the origin of the Loop I/NPS structure, including its position and physical size, has persisted for over $\sim$40 years and remains contentious (e.g. \citealt{Dickinson2018Galax,Planck2016XXV,Kataoka2018Galax,Shchekinov2018Galax}). 
Because of its huge angular size ($\sim 120\degree$, \citealt{Dickinson2018Galax}), many people consider it to be a nearby object within $\lesssim 1$ kpc, but the exact distance and geometry are still uncertain.
One method for deriving the distance of the Loop I/NPS is to analyze the alignment between {\black the polarization angles of} the optical starlight and the synchrotron {\black in the same direction}. These two angles should be highly correlated with each other if the synchrotron radiation comes from sources closer than the stars, since the starlight  polarization is {\black expected to be} generated by dust that traces the magnetic field, {\black which also produces the synchrotron}. Moreover, the thermal dust emission polarization also traces the magnetic field. \citet{Panopoulou2021ApJ} analyzed the polarization angles of these three tracers and derived upper limits on the Loop I/NPS distance, being all below  $\sim 150$ pc. It strongly favors that the Loop I, at least the part at $b>30\degree$, is a nearby source. 
Some authors suggested that it is a superbubble neighboring the Local Hot Bubble (LHB) where we {\black reside \citep{Cox1987ARA&A},} created by stellar winds and a series of supernova explosions happened in recent several million years in the Scorpius–Centaurus OB association that is $\sim 170$ pc away from the Sun
\citep{deGeus1992AA,Breitschwerdt2006AA}. 
The Geminga pulsar was produced in one of these supernova explosions \citep{Burke_Book2019}.
It was further pointed out that the Loop I/NPS superbubble is interacting with the LHB where we reside, and the interaction zone forms a dense HI ring (wall) between those two bubbles. The existence of the dense HI ring is supported by the sudden increase of HI column density at $\sim 40$ pc from Sun, indicated by the optical and UV absorption spectra of stars toward the center of the Loop I/NPS \citep{Centurion1991ApJ}; the coherence of X-ray absorption shadow and the HI column density distribution \citep{Egger1995A&A}, and the strengthened intensity of far-UV OVI emission line \citep{Sallmen2008ApJ}.

On the other hand, \citet{Reis2008AA} analyzed the color excess of more than 4000 stars up to 500 pc from the Sun, and \citet{Santos2011ApJ} analyzed the optical polarization of more than 800 stars. They found that the interface between the Loop I/NPS  
and the LHB is fragmented and distorted.
Such a discovery questions the notion of a coherent ring structure formed by the interaction of the Loop I/NPS with the LHB. Moreover, the analysis of X-ray absorption by the gas between the Loop I/NPS and the Sun indicates that the Loop I/NPS could be more distant, about $\sim 500 -1000$ pc away \citep{Iwan1980ApJ,Sofue2015MNRAS,Lallement2018A&A}. Faraday rotation depolarization and Faraday tomography methods also favour a distance of several hundred pc for the Loop I/NPS; see reviews  \citet{Dickinson2018Galax} and \citet{Lallement2023CRPhy}, and references therein.

Some people instead propose that the Loop I/NPS may originate from the Galactic Center (GC) \citep{Sofue1974PASJ,Sofue2000ApJ}, based on its directional proximity and possible association with the eROSITA bubble in the X-ray band, and with the Fermi Bubble in the gamma-ray band \citep{SuMeng2010ApJ,Predehl2020Natur}. 
Both the eROSITA bubble and the Fermi bubble are  
roughly symmetric with respect to the Galactic plane, strongly favoring the GC hypothesis \citep{Sofue1977A&A,Sofue2000ApJ,Sofue2016MNRAS}. In such a scenario, since the structure is located at $\sim 8$ kpc away, it must have considerable physical size \citep{Sarkar2019MNRAS,Sofue1977A&A,Sofue1994ApJ}. Using morphology analysis, \citet{Liu2024ApJ} also proposed that the NPS/eROSITA bubble is a distant and giant structure close to the GC.
\citet{Sofue2017PASJ} measured the HI gas density in the inner Milky Way, and found that there is an HI hole around the GC, and the hole has a crater-shaped wall that coincides with the Loop I/NPS. \citet{Zhang2024NatAs} investigated the distance to the polarized radio counterparts, including the Loop I/NPS, of the eROSITA bubbles by analyzing the multi-wavelength Faraday rotation depolarization. 
The wavelength-dependent depolarization reveals that the magneto-ionic medium responsible for the observed depolarization must extend up to a distance of 5 kpc. This finding suggests a connection between the polarized Loop I/NPS and outflows originated from the GC. The GC origin hypothesis is also supported by the fact that the measured H$\alpha$-to-1.4 GHz radio intensity ratio for the Loop I/NPS is two orders of magnitude smaller than the typical shell-type SNRs \citep{Sofue2023MNRAS}. The North-South asymmetry is explained by the lower density of the medium in the southern hemisphere. Expansion in a low-density medium results in weak shock acceleration efficiency, and hence weak radiation \citep{Lallement2023CRPhy,Sarkar2019MNRAS}.

In  the radio band, Loop I/NPS's emission is mainly the synchrotron radiation \citep{Borka2007MNRAS}.
There have been many models developed to explain the Loop I/NPS's morphology. In the supernova remnants (SNRs) scenario, the bright ridge originates from a compressed magnetic field and/or enhanced cosmic ray, i.e., the emissivity at the shell is larger than the interior.
The asymmetry of the loop is due to the inhomogeneous distribution of the gas (see \citealt{Dickinson2018Galax} and references therein). In the ultra-long wavelength band (wavelength $\gtrsim 10$ m or frequency $\lesssim 30$ MHz), particularly at frequencies $\lesssim 10$ MHz, the free-free absorption by diffuse electrons in the interstellar medium (ISM) becomes important and may change the observed morphology of the radio sky, including the Loop I/NPS structure \citep{Cong2021APJ,Cong2022ApJ}. The Loop I/NPS, at least the low Galactic latitude part, is expected to be darker if it is located at the GC, but will still be bright if it is a complex of nearby SNRs. \citet{Cong2021APJ} pointed out that the ultra-long wavelength observations have the potential to solve the problems of Loop I/NPS's origin. In this paper, we explicitly investigate this problem. We develop phenomenological models to predict the observational features of the Loop I/NPS for both the SNRs scenario and the GC scenario, in the ultra-long wavelength band.  The outline of this paper is as follows. In Sec. \ref{sec:methods}, we present the emissivity model for the Galactic disk and the Loop I/NPS structure in these two scenarios. Then we develop an electron density model for the Loop I/NPS.
{\black The main results and discussion on some uncertainties are presented in Sec. \ref{sec:result}.}
We summarize the results in Sec. \ref{sec:summary}  

\section{Methods}\label{sec:methods}

In this work, we model the Galactic emissivity as comprising of a smooth disk component, including both {\black the} thin and thick disks
\footnote{ It is worth noting that this is not the same concept as the optical thin disk and thick disk; these thin and thick disks describe the diffuse components {\black in radio bands}, primarily originating from synchrotron radiation.}, and a Loop I/NPS structure. Model parameters are derived by reproducing the observed Haslam 408 MHz all-sky map \citep{Haslam1982A&A,Remazeilles2015MNRAS}, and extrapolated to lower frequencies by a power-law formula. 

\subsection{The disk emissivity}

{\black  
We assume that the Galactic disk synchrotron emissivity is composed of a thin disk component and a thick disk component. The thin disk is geometrically thin but dominates the Galactic plane and around. It is associated with star formation, regular magnetic field and SNRs around Galactic spiral arms. The thick disk is more extended in the vertical direction and dominates the emission at high Galactic latitudes. It reflects the transport and diffuse processes of cosmic rays, and both random and regular Galactic halo magnetic fields (e.g. \citealt{Mertsch2013JCAP,Jansson2012ApJ,Sun2008A&A}). 
As for the radiation from the Loop I/NPS and its vicinity, the contribution from both high latitudes and low latitudes is important, it is necessary to model both of them. Although this two-disk scenario has not yet been confirmed in observations, we find that it fits the 408 MHz sky map better than the one-disk model adopted in \citet{Cong2021APJ}. 
}

{\black The disk emissivity can be written as
\begin{align}
\epsilon_{\rm disk}(\nu|R,Z)
=\sum_{i}  A_{i} \exp \left\{-{ b(\alpha_i)}\left[ \left(\frac{R}{R_i}\right)^{\frac{1}{\alpha_i}} -1 \right]\right\} \text{sech}\left( \frac{Z}{Z_i}\right) \left( \frac{\nu}{\nu_*}\right)^{\beta_G}, 
\label{eq:emiss_disk}
\end{align}
where  $i \in [{\rm thin}, {\rm thick}]$. $R$ and $Z$ are cylindrical Galactic coordinates, where $R$ is the radial distance to the GC, while $Z$ is the vertical distance to the Galactic plane. Here $\nu$ is the frequency, and $\beta_G$ is the spectral index.} The radial dependence form of Eq. (\ref{eq:emiss_disk}) is the Sérsic profile \citep{sersic1968} that is widely used to fit the surface brightness distribution of elliptical galaxies, as well as the disk and bulge components of other galaxy types.
$A_i$ denotes the emissivity at the effective radius $R_i$, which encompasses half of the total intensity in $Z=0$ plane. The index $\alpha_i$ indicates the profile's curvature in the radial direction. {\black $Z_i$ is the height scale. Finally, the function $b(\alpha)$ is a dimensionless scale factor. It is the solution to
\begin{equation}
\Gamma(2\alpha,b)=2\int_0^b t^{2\alpha-1}e^{-t}dt,
\end{equation}
where $\Gamma$ is the Gamma Function. We use the following approximation \citep{Ciotti_Bertin1999,MacArthur2003ApJ}
\begin{equation}
b(\alpha)=
\begin{cases}
0.01945-0.8902\alpha+10.95\alpha^2-19.67\alpha^3+13.43\alpha^4 &\alpha \le 0.36  \\
2\alpha - \frac{1}{3} + \frac{4}{405\alpha} + \frac{46}{25515\alpha^2} + \frac{131}{1148175\alpha^3} + \frac{2194697}{30690717750\alpha^4} & \alpha >0.36.
\end{cases}
\end{equation}
}

The frequency-independent parameters $A_i$, $R_i$, $\alpha_i$, and $Z_i$, {\black together with the parameters in the Loop I/NPS emissivity model to be introduced in Sec. \ref{sec:LoopI-emissivity}}, are obtained by fitting the {\black Haslam 408 MHz map \citep{Haslam1982A&A,Remazeilles2015MNRAS}, for which the free-free emission and the isotropic extragalactic radiation are pre-subtracted.}
%{\black using the sum of the disk and the Loop I/NPS radiation. We will introduce the Loop I/NPS emissivity model in Sec. \ref{sec:LoopI-emissivity}, and
{\black The fitted parameters of both the disk component and the Loop I/NPS component are given in Tab. \ref{tab:free_params_and_LoopI_params}. 
When pre-processing the Haslam 408 MHz map, the free-free emission can be derived from the H$\alpha$ emission line. However, there are uncertainties induced by dust absorption and scattering at the Galactic plane and in some dense HII regions (e.g. \citealt{Dickinson2003MNRAS}). 
We derive the frequency-dependent free-free sky maps from the emission measure (EM) and electron temperature sky maps provided by \citet{Plank2016A&A}, taking into account the free-free self-absorption. The formulae are listed in Eq. (14) of \citet{Cong2021APJ}. The free-free sky map at 408 MHz and the global spectrum are shown in Fig. 2 of that paper. At 408 MHz, for high Galactic latitude regions with $|b|\gtrsim10\degree$, free-free emission only contributes $\sim$1\% to the total intensity. Due to the shallower spectrum index compared with synchrotron ($\beta_{\rm ff}\sim -2.1$ for free-free emission vs. $\beta_G\sim -2.5$ for synchrotron), this contribution becomes even negligible below $\sim 10$ MHz. Free-free emission is highly concentrated on the Galactic plane. In some particularly dense HII regions (for example, the dense HII regions in Orion-Eridanus Superbubble, Perseus-Taurus Supershell, and Ophiuchus Superbubble), the free-free model would be inaccurate. It may underestimate the free-free emission and absorption. However, the present work focuses on the Loop I/NPS and regions around it; this is large-scale morphology, and such flaws will not change our conclusions. We model the extragalactic background as isotropic radiation,
$T_{\rm eg}=1.2(\frac{\nu}{{\rm 1GHz}})^{-2.58}$, derived from ARCADE-2 observations \citep{Seiffert2011ApJ}. 
}

{\black Throughout this paper, we use a constant spectrum index, $\beta_G=-2.51$, for Galactic synchrotron emissivity  \citep{Cong2021APJ}. This value is derived by fitting the observed sky maps at 10 frequencies from 35 MHz to 408 MHz \citep{Dowell2017MNRAS,Guzman2011A&A,Haslam1982A&A,Remazeilles2015MNRAS}.
}

\subsection{The Loop I/NPS emissivity}\label{sec:LoopI-emissivity}

{\black
No matter whether the Loop I/NPS is a superbubble created by stellar winds or supernova explosion shocks, or is a giant bubble located at the GC and created by outflows associated with star formation/black hole activities, the ambient medium is swept and compressed, and it must finally form a cavity with a dense shell. The shell could be incomplete or asymmetric if the medium is inhomogeneous. Motivated by this,
} 
we begin by modeling the Loop I/NPS emissivity as a sphere with a thin and dense spherical shell, which we then trim to reproduce the observed loop morphology. 
{\black 
Similar scenarios have been extensively employed in studies of radio loop morphology (e.g. \citealt{Wolleben2007ApJ,Mertsch2013JCAP,Mou2023A&A}).  
}
Let the radius of the sphere be $r_{\rm L}$ and the thickness of the shell be $\Delta r_{\rm s}$, then the emissivity of the Loop I/NPS can be written as:   
\begin{equation}
\epsilon_{\rm L}(r)=\begin{cases}
\epsilon_{\rm i}~~~~&r<r_{\rm L}-  \Delta r_{\rm s} \\
\epsilon_{\rm s}~~~&r_{\rm L}-\Delta r_{\rm s}\le r\le  r_{\rm L}  \\
0  & r > r_{\rm L},
\end{cases}
\end{equation}
where $\epsilon_{\rm i}$ is the emissivity of the sphere's interior, $\epsilon_{\rm s}$ is the emissivity of the shell, and $r$ is the distance to the center of the Loop I/NPS sphere. 
Given that $\epsilon_s \gg \epsilon_i$, the projection of the sphere onto the celestial sphere results in a limb-brightened loop.
{\black
By adjusting the sphere's radius, the shell thickness, and its emissivity, and by selecting which part of the shell to trim, the model can accurately reproduce the observed bright Loop I/NPS arc.
}

In this study, we examine two potential scenarios for the origin of Loop I/NPS: the shell-like SNRs model, which attributes {\black the} Loop I/NPS to supernova remnants near the Sun; and the GC model, which places {\black the Loop I/NPS close to the center of the Milky Way}, akin to the Fermi Bubble.
For both models, the angular size of {\black the} Loop I/NPS is approximately $\sim 116\degree$ \citep{Mertsch2013JCAP}, {\black implying a large difference in its physical size}.

\subsubsection{The shell-like SNRs model}

In this model, we adopt the direction (the Galactic longitude and latitude) of the Loop I/NPS center and its distance to {\black the} Sun given in \citet{Berkhuijsen1971A&A,Mertsch2013JCAP}, 
{\black i.e., the center of the Loop I/NPS sphere is at Galactic 
Cartesian coordinates ($X_L=-8.3$, $Y_L=-0.12$, $Y_L=0.07$) kpc.} The observed morphology of the Loop I/NPS exhibits irregular characteristics.
This is likely a result of the interaction between supernova remnants and a joint envelope formed by the stellar winds of the Sco-Cen OB association or the LHB \citep{Egger1995A&A,Vidal2015MNRAS}. Thus, it is necessary to eliminate certain portions of the emissivity sphere to reproduce the observed morphology accurately.

We first remove the portion of the Loop I/NPS below the Galactic plane, $Z<0$, to address the north-south asymmetry.
Then we create a suppositional 3D ellipse centred at Galactic {\black 
Cartesian coordinates $(X=-8.522, Y=-0.089, Z=0.029)$ kpc} with axes of (2$r_{\rm L}$, 0.4$r_{\rm L}$, 2$r_{\rm L}$), and remove the volume in the Loop I/NPS overlaps with this ellipse.
This is motivated by the east-west asymmetry in the observed morphology.
Following this processing, we successfully produce the morphological features of the Loop I/NPS as observed in the Haslam 408 MHz map.

\subsubsection{The Galactic Center model}\label{sec:LoopI-GC}

In the GC model, the Loop I/NPS is considered as the post-shock medium \citep{Mou2023A&A}. 
{\black The center of the sphere is at Galactic 
Cartesian coordinates ($X_L=0.0$, $Y_L=1.5$, $Y_L=5.0$) kpc.} It consistently aligns with the shape of the Fermi Bubble and the X-ray Bubble \citep{Mou2023NatCo}.
Similar to the shell-like SNRs model, here we also remove the portion where $Z < 0$. 
{\black However, in the GC model, the $Z<0$ portion is a small fraction compared to its northern counterpart because now the center of the Loop I/NPS is well above the Galactic plane.
}
Then we create a suppositional sphere in 3D space centered at {\black Galactic 
Cartesian coordinates} $(X_{\rm L}=-5.262, Y_{\rm L}=-5.608, Z_{\rm L}=1.142)$ $\rm kpc$ with a radius of $r_{\rm L}$ and eliminate the overlap between the Loop I/NPS and the sphere to address the east-west asymmetry.

\begin{table}[h!]  
\centering  
\caption{{\black The frequency-independent parameters for the thin and thick disks and the Loop I/NPS emissivity, in the SNRs and GC models for the Loop I/NPS, derived by fitting the observed 408 MHz Haslam sky map after extrating the free-free emission and the extragalactic background. It is natural that in both the SNRs and GC models of the Loop I/NPS, the parameters for disks are close to each other. 
$X_{\rm L}$, $Y_{\rm L}$, and $Z_{\rm L}$ are the assigned Galactocentric Cartesian coordinates of the Loop I/NPS center.  
}}  
\begin{tabular}{l|cc|cc}  
\hline  
\hline  
Disk parameters                & \multicolumn{2}{c|}{in the SNRs model}             & \multicolumn{2}{c}{in the GC model} \\  
\hline  
$A_{\rm thick}$ (K/kpc)             & \multicolumn{2}{c|}{1.65}                         & \multicolumn{2}{c}{1.73}                       \\  
$R_{\rm thick}$ (kpc)               & \multicolumn{2}{c|}{5.67}                         & \multicolumn{2}{c}{5.73}                       \\  
$\alpha_{\rm thick}$                & \multicolumn{2}{c|}{0.15}                         & \multicolumn{2}{c}{0.14}                       \\  
$Z_{\rm thick}$ (kpc)               & \multicolumn{2}{c|}{3.04}                         & \multicolumn{2}{c}{2.80}                        \\  
$A_{\rm thin}$ (K/kpc)                          & \multicolumn{2}{c|}{5.01}              & \multicolumn{2}{c}{5.0}           \\  
$R_{\rm thin}$ (kpc)                            & \multicolumn{2}{c|}{7.82}                & \multicolumn{2}{c}{8.30}           \\  
$\alpha_{\rm thin}$                             & \multicolumn{2}{c|}{1.94}                & \multicolumn{2}{c}{2.09}           \\  
$Z_{\rm thin}$ (kpc)                            & \multicolumn{2}{c|}{0.3}                 & \multicolumn{2}{c}{0.3}            \\ 
\hline   
Loop I/NPS parameters               & \multicolumn{2}{c|}{}             & \multicolumn{2}{c}{} \\
\hline  
$X_{\rm L}$ (kpc)       & \multicolumn{2}{c|}{-8.3}                & \multicolumn{2}{c}{0.0}                            \\  
$Y_{\rm L}$ (kpc)       & \multicolumn{2}{c|}{-0.12}                & \multicolumn{2}{c}{-1.5}                             \\  
$Z_{\rm L}$ (kpc)       & \multicolumn{2}{c|}{0.07}                & \multicolumn{2}{c}{5.0}                             \\  
$r_{\rm L}$ (kpc)       & \multicolumn{2}{c|}{0.22}                & \multicolumn{2}{c}{7.8}                            \\  
$\Delta r_{\rm s}$ (kpc) & \multicolumn{2}{c|}{0.02}                & \multicolumn{2}{c}{2.5}                             \\  
$\epsilon_{\rm s}$ (K/kpc) & \multicolumn{2}{c|}{129.68}                & \multicolumn{2}{c}{1.37}                      \\  
$\epsilon_{\rm i}$ (K/kpc) & \multicolumn{2}{c|}{0.001}                & \multicolumn{2}{c}{0.26}                       \\  
\hline  
\hline  
\end{tabular}  
\label{tab:free_params_and_LoopI_params}  
\end{table}

Finally, the total emissivity of the Milky Way is the sum of emissivities of the thin disk, the thick disk, and the Loop I/NPS,
\begin{equation}
    \epsilon_{\rm MW}(\nu|R,Z) = \epsilon_{\rm disk}(\nu|R,Z) + \epsilon_{\rm L}(\nu|R,Z).
    \label{eq:emissivity}
\end{equation}
{\black We then fit the observed 408 MHz sky map (free-free emission and extragalactic background subtracted) with Eq. (\ref{eq:emissivity}) (the coordinates of the Loop I/NPS center are specified in each model). 
The parameters derived for the thin and thick disks, as well as for the Loop I/NPS, are listed in Tab. \ref{tab:free_params_and_LoopI_params}. 
When the adopted Loop I/NPS model switches from SNRs to GC, we always obtain consistent parameters for the disk. It is interesting to note that, our $Z_{\rm thin}$ is roughly consistent with the observed optical thin disk which contains younger stars, gas, and dust with a scale height of $\sim 120-300$ pc; although our $Z_{\rm thick}$ is a bit larger than the scale height of optical thick disk (typically $\sim 0.5-1.9$ kpc) that contains elder stars (see \citealt{Vieira2023Galax} and references therein).
}

\begin{figure*}[t]
	\centering
	{\includegraphics[width=0.45\textwidth]{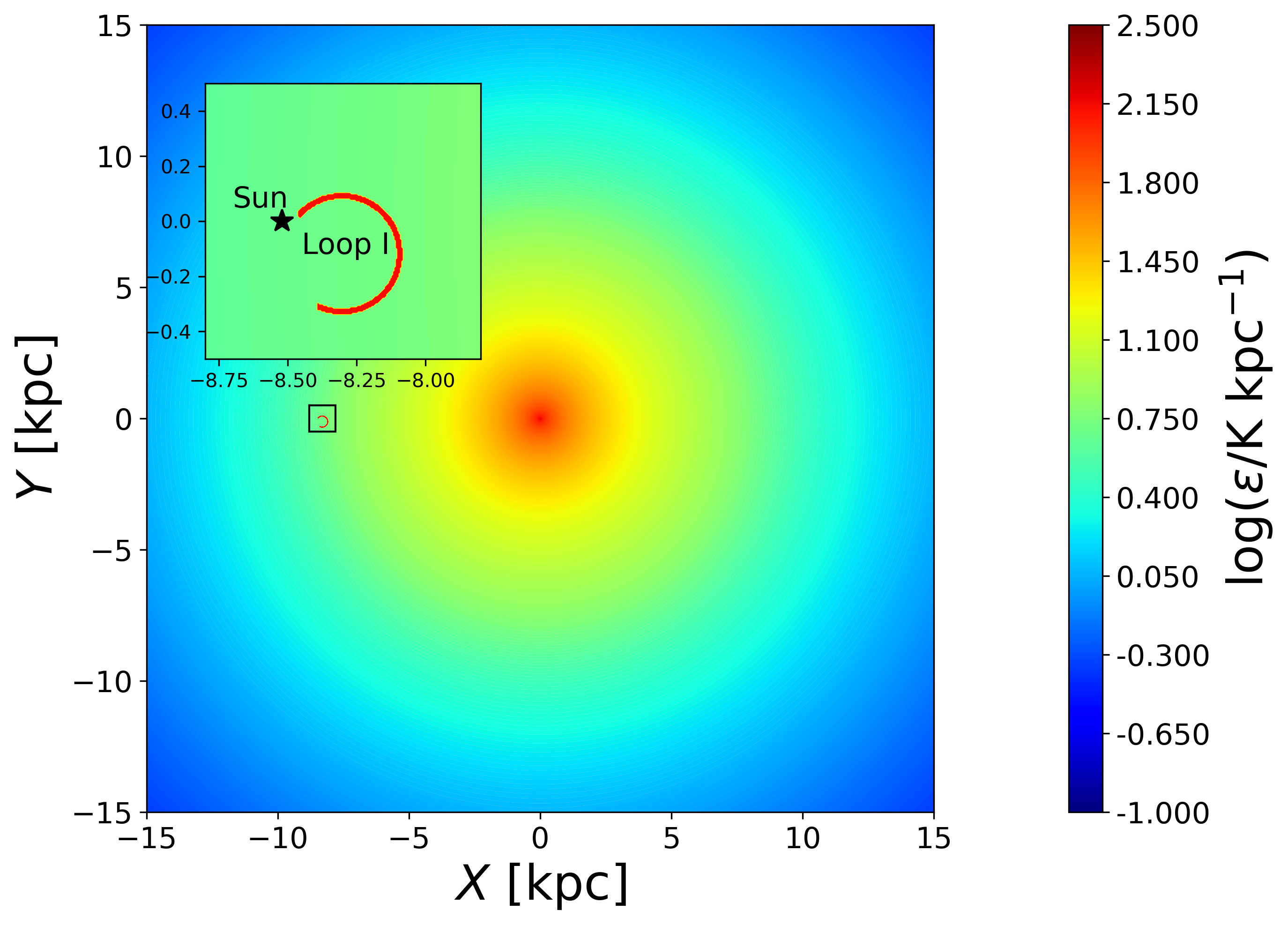}}
	 {\includegraphics[width=0.45\textwidth]{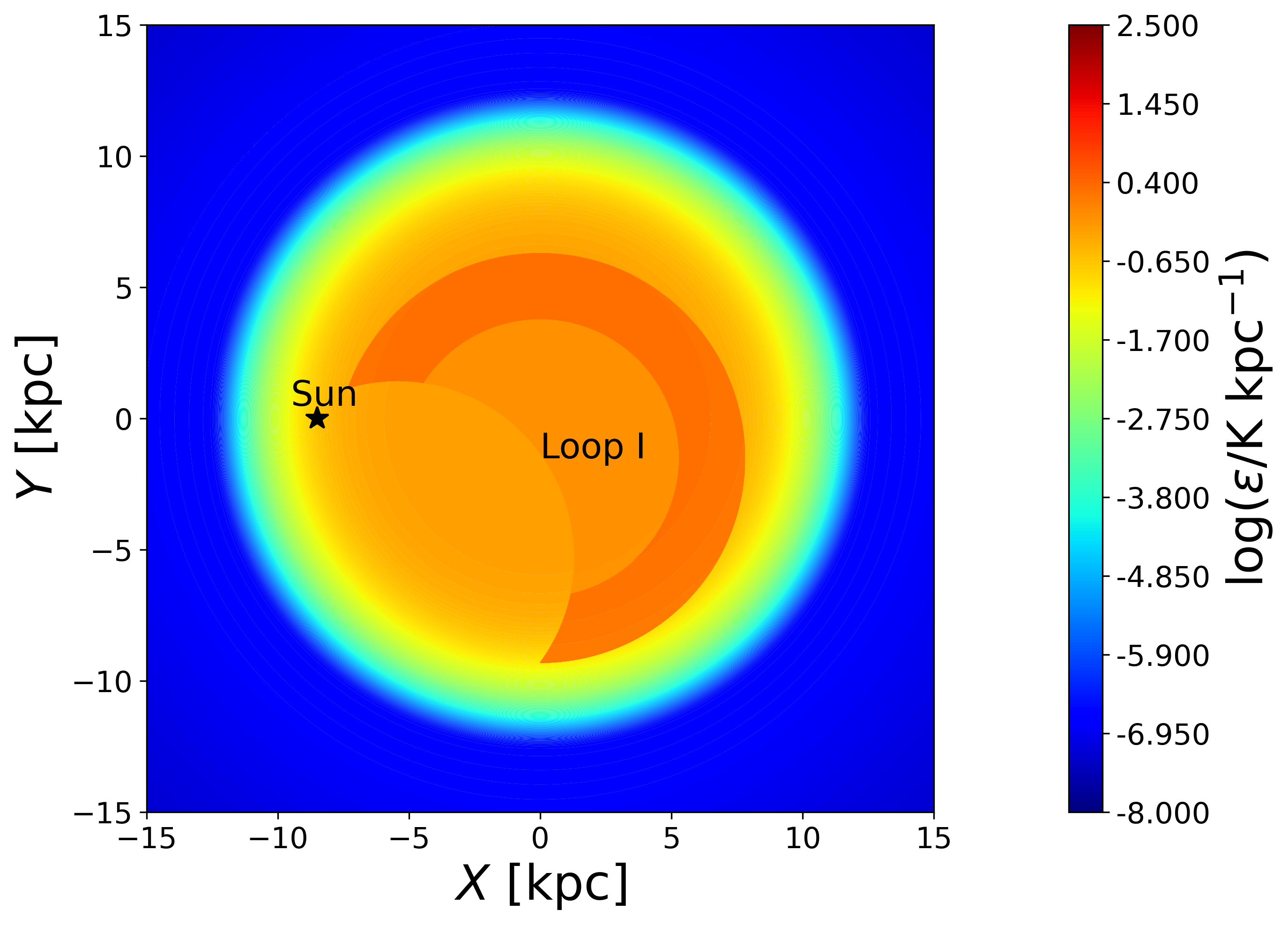}}
	 
	 {\includegraphics[width=0.45\textwidth]{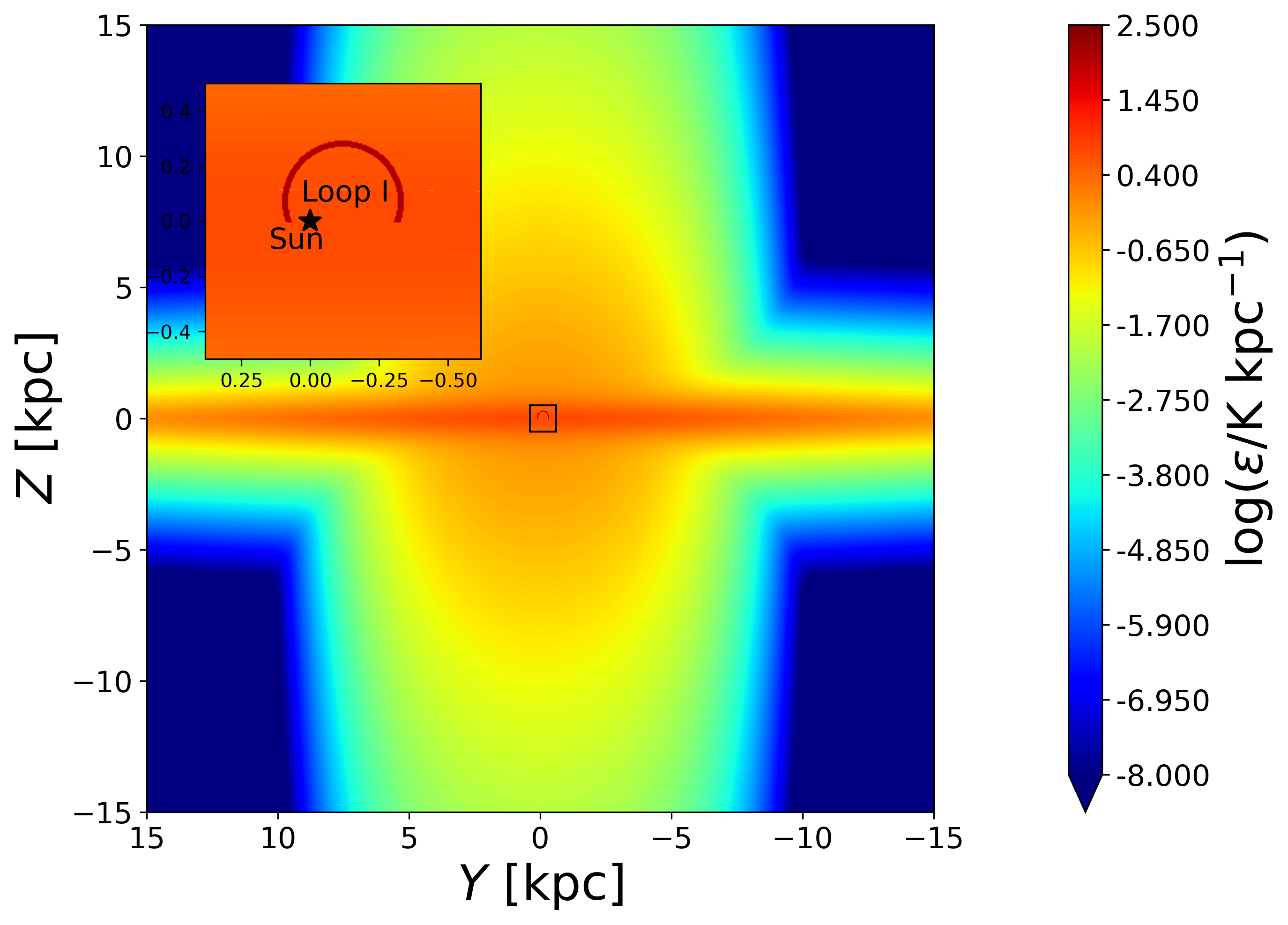}}
	 {\includegraphics[width=0.45\textwidth]{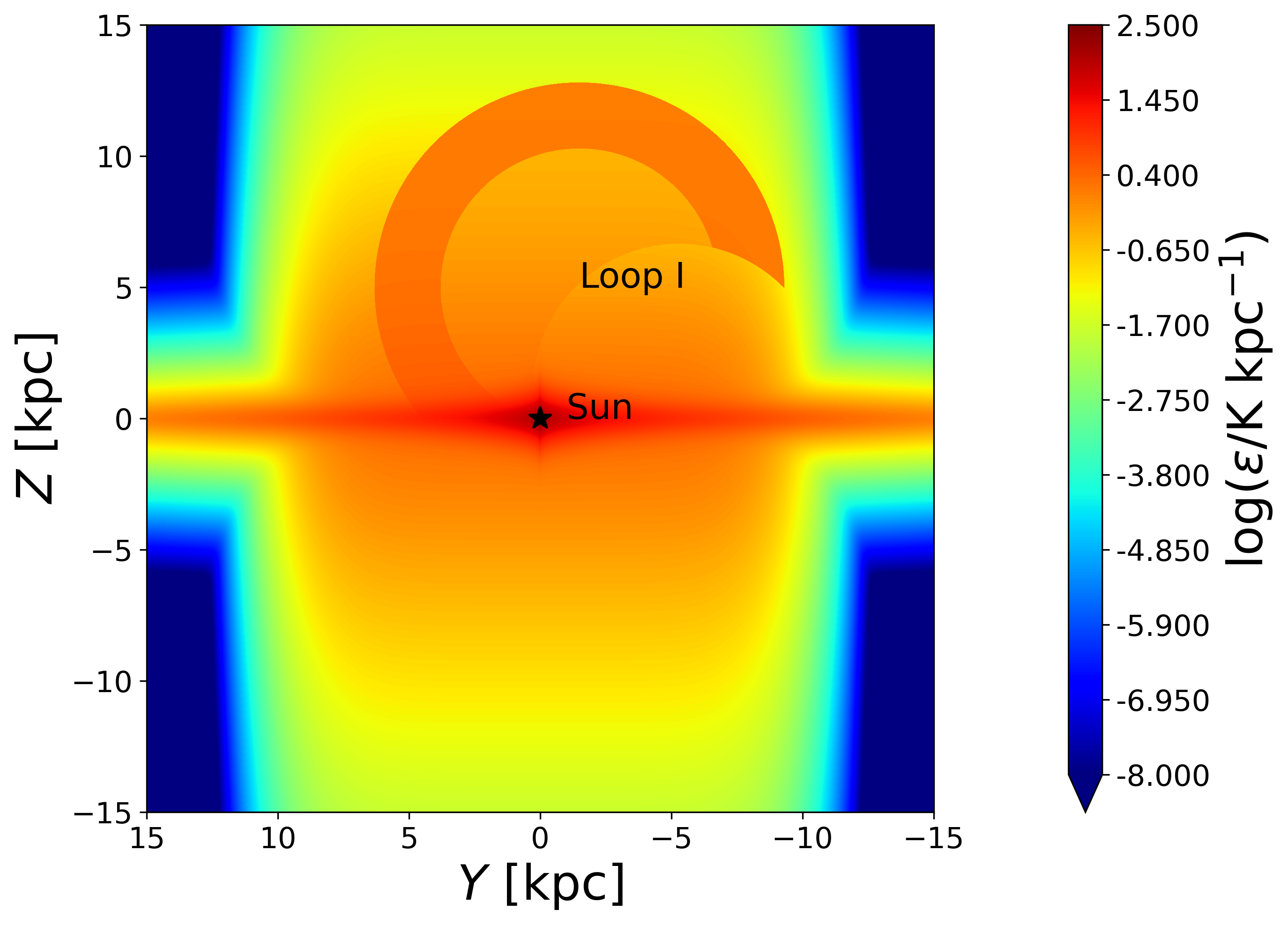}}
	\caption{
	{\black 
        Emissivity distributions at 408 MHz in Galactocentric coordinates for the two models of the Loop I/NPS: shell-like SNRs (left) and GC (right). The slices are in the $X-Y$ plane (face-on, top), and the $Y-Z$ plane (edge-on, bottom), respectively.
        }
	}\label{fig:emi_2D}
\end{figure*}

\begin{figure*}[t]
	\centering
	 {\includegraphics[width=0.4\textwidth]{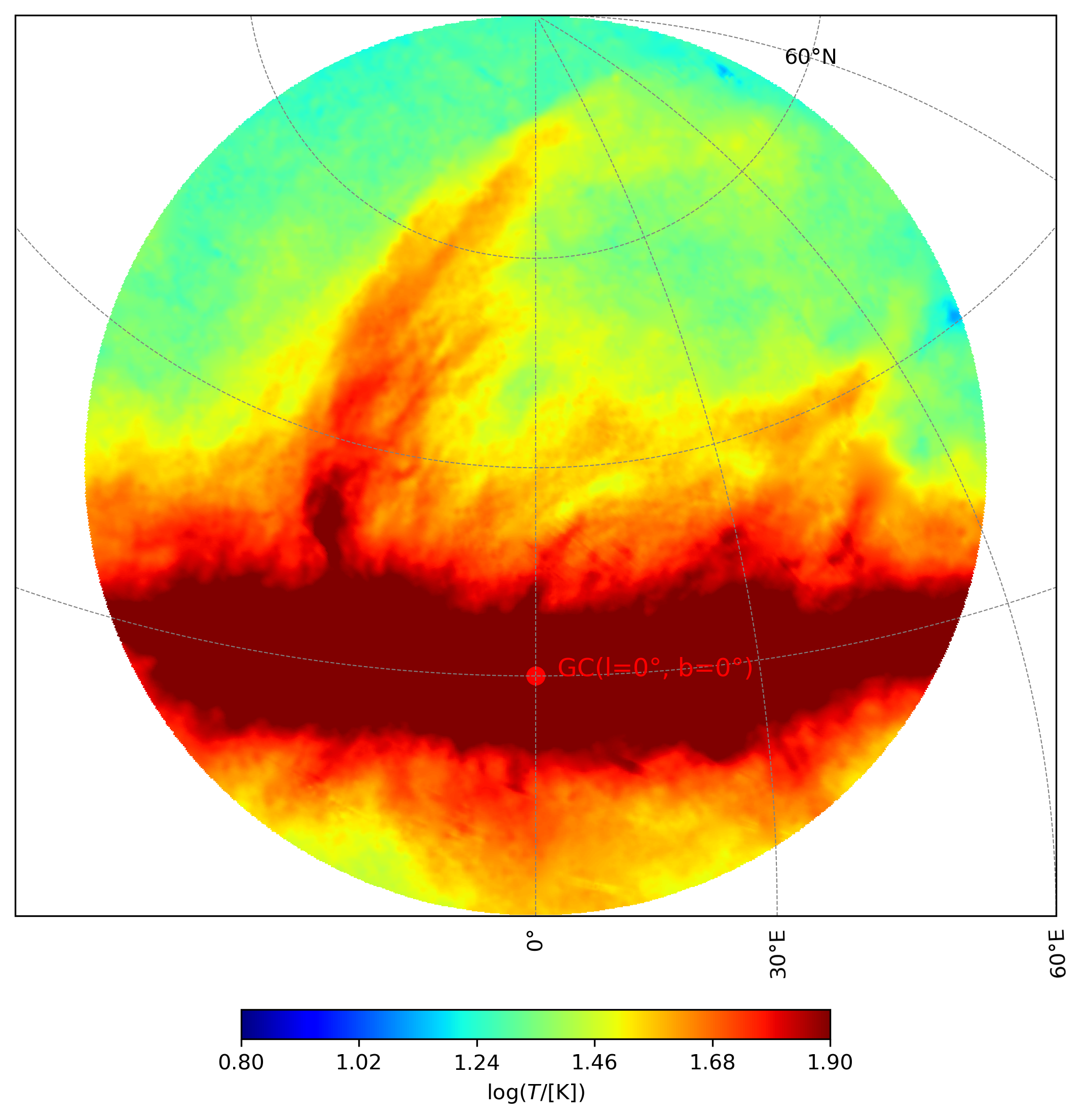}}	 {\includegraphics[width=0.4\textwidth]{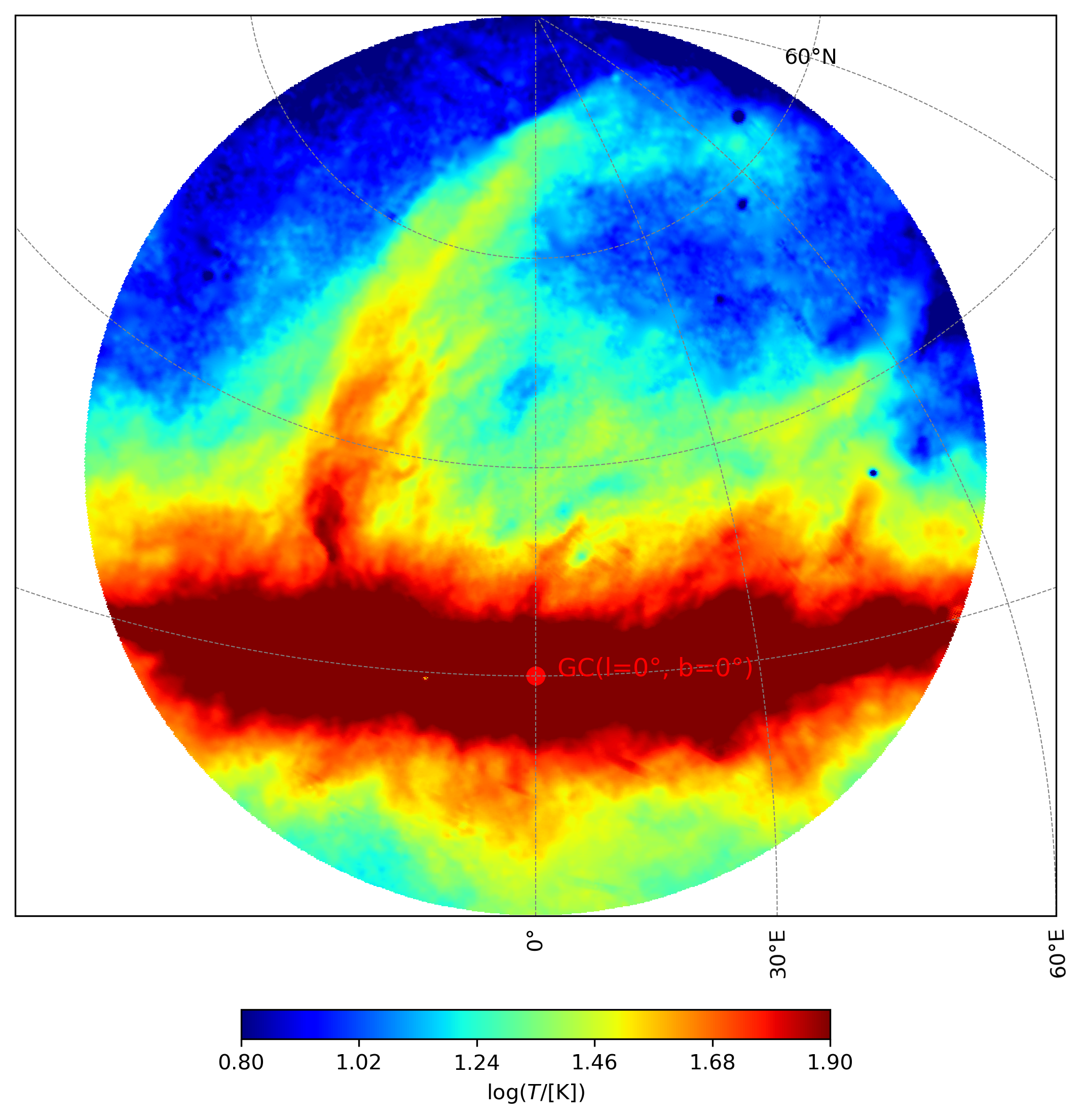}}     {\includegraphics[width=0.4\textwidth]{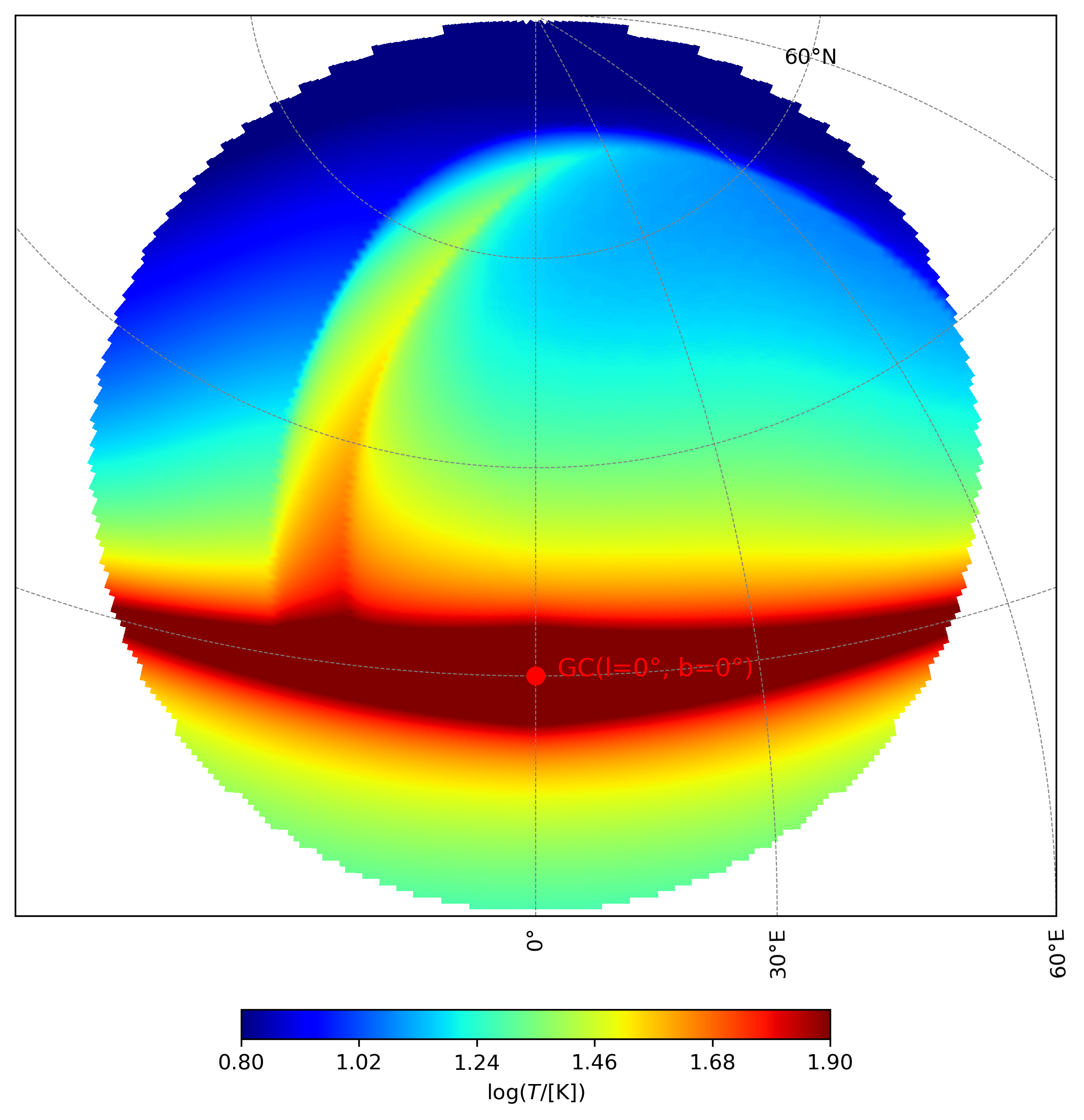}}     {\includegraphics[width=0.4\textwidth]{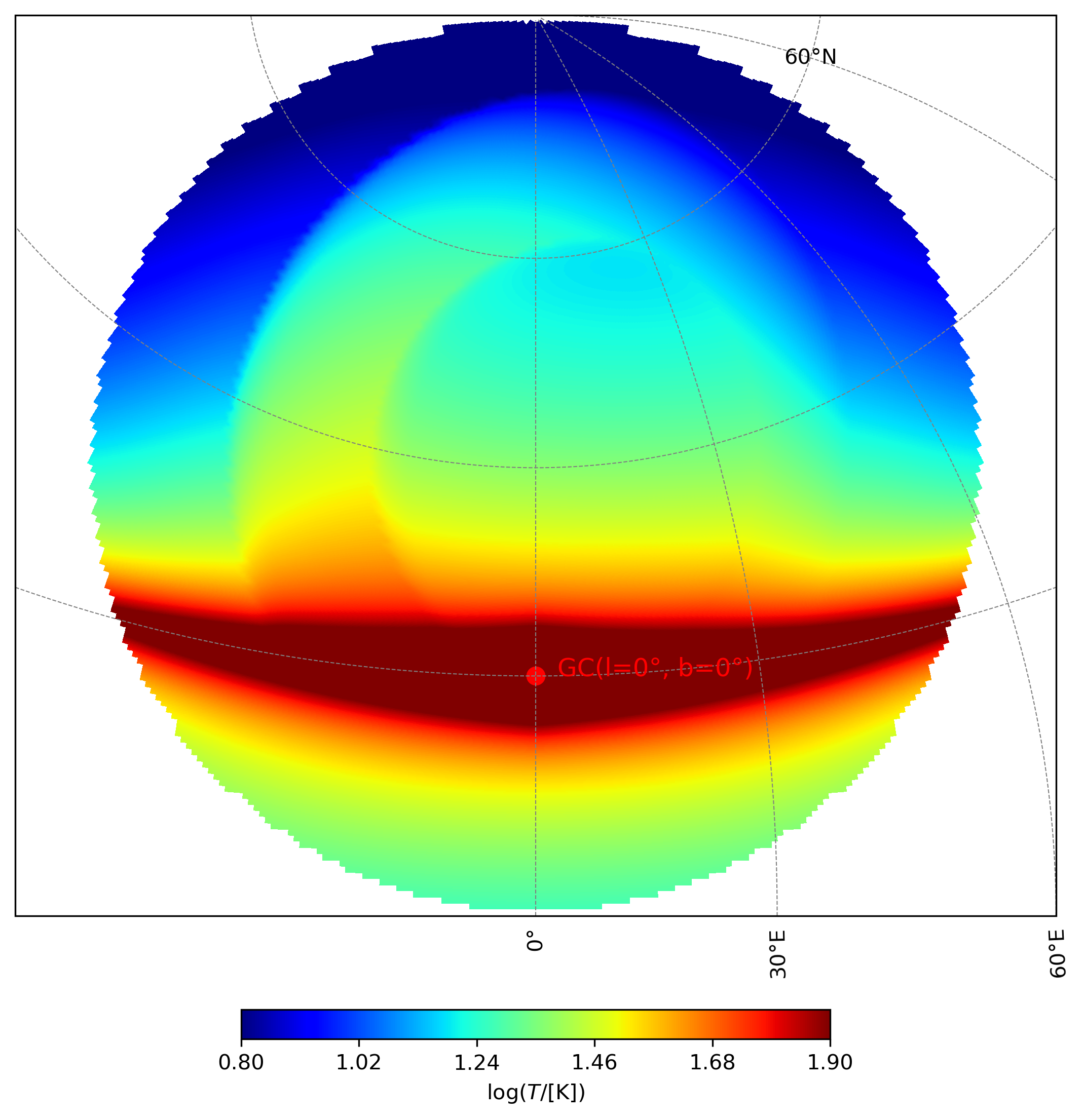}}
     {\includegraphics[width=0.4\textwidth]{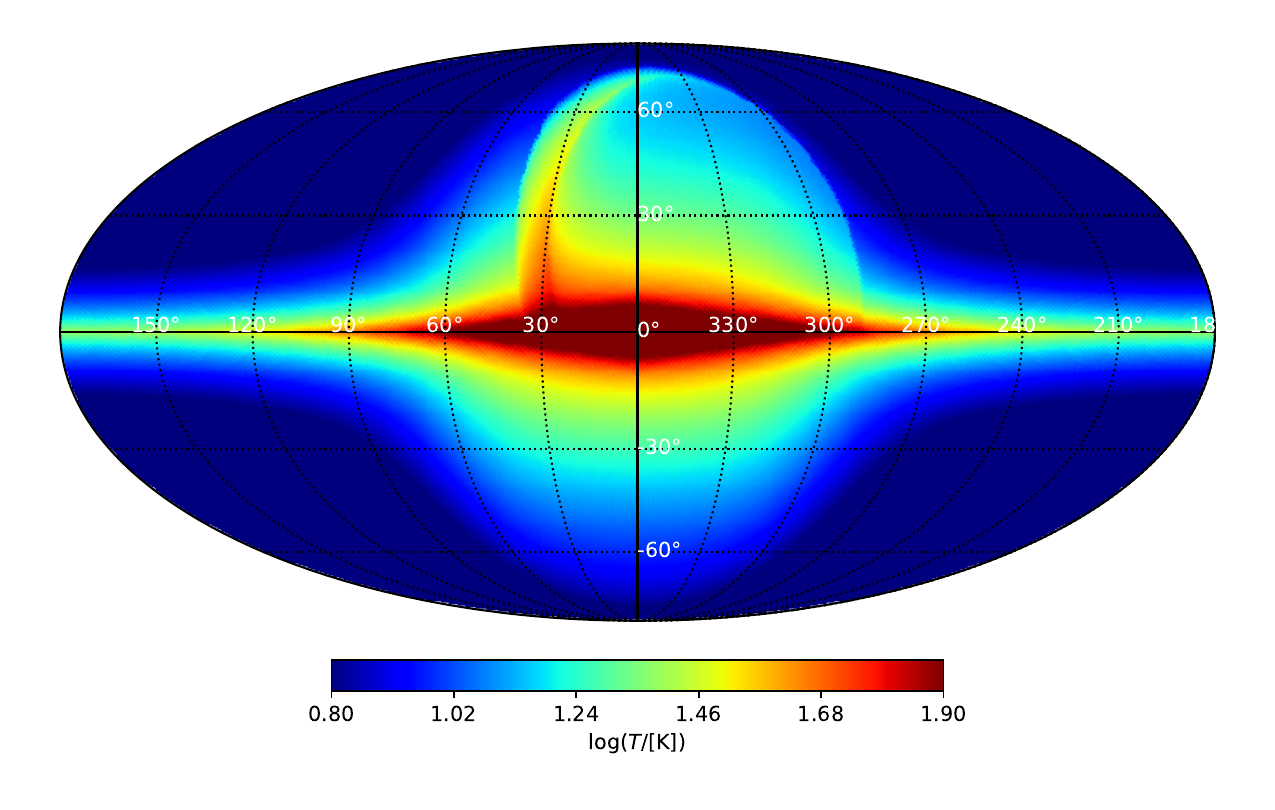}}
     {\includegraphics[width=0.4\textwidth]{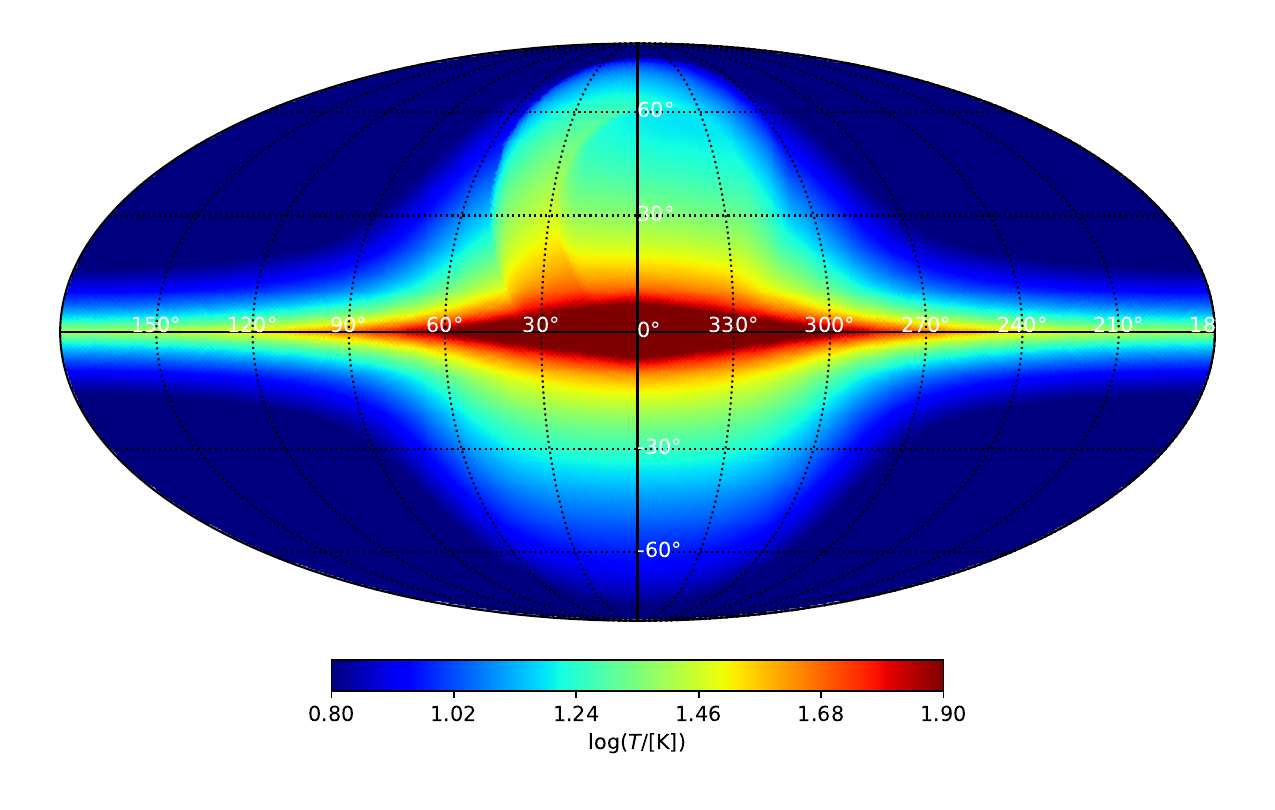}}
        \caption{\black 
{\it Top left}: The Loop I/NPS structure and its vicinity in raw Haslam 408 MHz sky map, viewed in stereographic projection. {\it Top right}: Similar to top left panel, however the free-free emission and extragalactic background have been subtracted.
{\it Middle left}: Our predicted Loop I/NPS and its vicinity at 408 MHz in stereographic projection, for the SNRs model. {\it Middle right}: Similar to the middle left panel, but the GC model for Loop I/NPS is adopted.
{\it Bottom left}: The predicted full sky map at 408 MHz, including both the disk and the Loop I/NPS, in Mollweide projection, for the SNRs model.  
{Bottom right}: Similar to the bottom left panel, but for the GC model.
All predicted sky maps do not include free-free emission and extragalactic background. The maps are displayed in Galactic coordinates, and the color scale uses a logarithmic stretch to better display the full model, particularly for regions away from the Galactic plane.   }\label{fig:408MHz_model_and_true}
\end{figure*}

Fig. \ref{fig:emi_2D} shows the emissivity distribution slices of the Milky Way, encompassing both the disk and the Loop I/NPS in the two models.
We obtain the sky map by integrating this emissivity along each line-of-sight and show the results at 408 MHz in Fig. \ref{fig:408MHz_model_and_true}. 
Our emissivity models replicate the morphology of the observed sky map very well.

\subsection{The electron density  of the Loop I/NPS}\label{subsec:electron_of_loopI}

{\black In order to model the free-free absorption by electrons at the ultra-long wavelengths, we adopt the}
{\tt NE2001} model for Galactic free electron distribution \citep{ne2001_I,ne2001_II}. It comprises five components: the {\black smooth component comprising the} thin and thick disks {\black and five} spiral arms, the Galactic Center, the local interstellar medium {\black including a few nearby voids}, {\black a list of} known dense clumps, and {\black a list of more distant} voids. In the {\tt NE2001} model, the Loop I/NPS is {\black one of the nearby voids in the local interstellar medium, manifesting} a hemisphere void above the Galactic plane, with a thin shell.  
In this work, we improve the Loop I/NPS structure in the electron density model. We assume that the geometry of the Loop I/NPS is the same as the emissivity model, but derive the electron densities from {\tt HaloSat} X-ray observations \citep{LaRocca2020ApJ}.

{\tt HaloSat} is a satellite designed to detect diffuse X-ray emissions in the range 0.4 - 7.0 keV \citep{Kaaret2019ApJ,LaRocca2020JATIS}. It has a full response field-of-view of \(10\degree\), which makes it a useful tool for probing the thermal X-ray emission from the Loop I/NPS free electrons. 
{\tt HaloSat} provides 14 pointing directions for the Loop I/NPS region. \cite{LaRocca2020ApJ} proposed that the Loop I/NPS consists of a two-phase plasma: a cool component with an average energy of 0.087 keV (temperature $\sim 1.01\times 10^6$ K) and a hot component with an average energy of 0.274 keV (temperature $\sim 3.18\times 10^6$ K) \footnote{\citet{LaRocca2020ApJ} supposes that the Loop I/NPS fully contribute to the fitted hot component, and primarily contribute to the fitted cool component.}.
Both of them are optically thin and in ionization equilibrium.
They fitted the EM values of the two components of the Loop I/NPS for the 14 pointing directions according to the observed X-ray spectra. The value of EM decreases with the increase in angle relative to the GC. 
We sum the EM for the cool and hot components to obtain the full EM for the Loop I/NPS. In principle, if the Loop I/NPS is an unknown structure at the GC, the observed EM is from both the Loop I/NPS and the ISM in-between. However, since the emissivity of the Loop I/NPS dominates over the intervening ISM in the radio sky, here we naively assume the measured EM is also mainly from the Loop I/NPS.

For the $i$-th line-of-sight, the EM of the Loop I/NPS is 
\begin{equation}
{\rm EM}_i=\sum_{j=1}^{N_i} \frac{1}{f_e} n^2_{e,j}\Delta s_{i,j}, 
\end{equation}
where this line-of-sight passes $N_j$ segments in the Loop I/NPS. For the $j$-th segment, the electron density is $n_{e,j}$, and the path length is $\Delta s_{i,j}$. $f_e$ is a filling factor and we use $f_e=0.5$ \citep{LaRocca2020ApJ}.

We divide the spherical coordinates $\theta$ of the Loop I/NPS-centered frame sphere shell into 20 parts and $\phi$ coordinates into 4 parts, {\black where $\theta$ is the polar angle and $\phi$ is the azimuthal angle in the regular spherical coordinate system.} The interior of the sphere is a single part with uniform density. We then perform Markov Chain Monte Carlo (MCMC, \citealt{emceee2013}) 
fitting for all the 14 line-of-sights in {\tt HaloSat} observations, to find the electron density for each segment. For the segment without any line-of-sight passing, the electron density is assigned by interpolation from the nearest segment.

\begin{figure*}[t]
	\centering
	 {\includegraphics[width=0.45\textwidth]{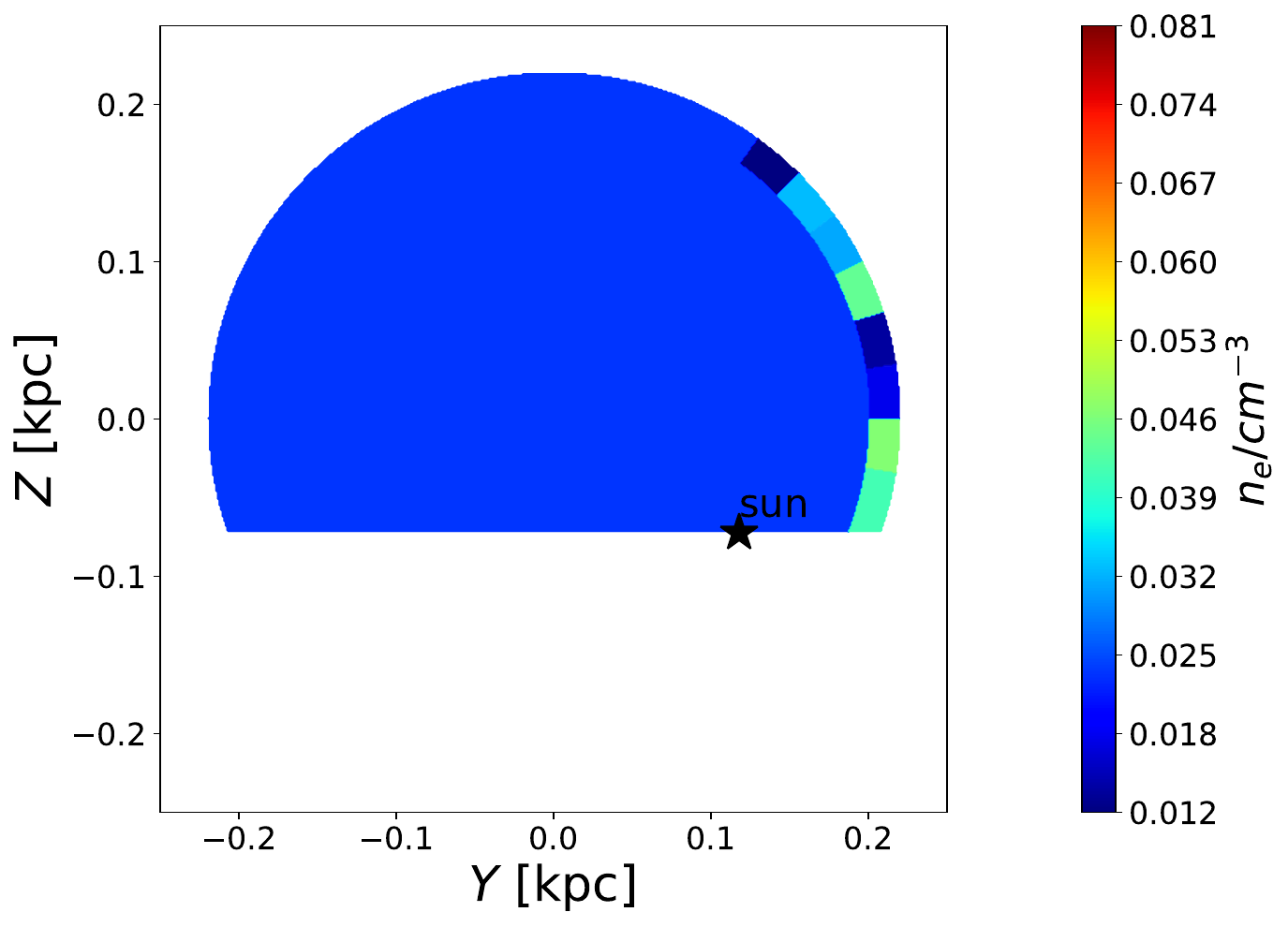}}
	  {\includegraphics[width=0.45\textwidth]{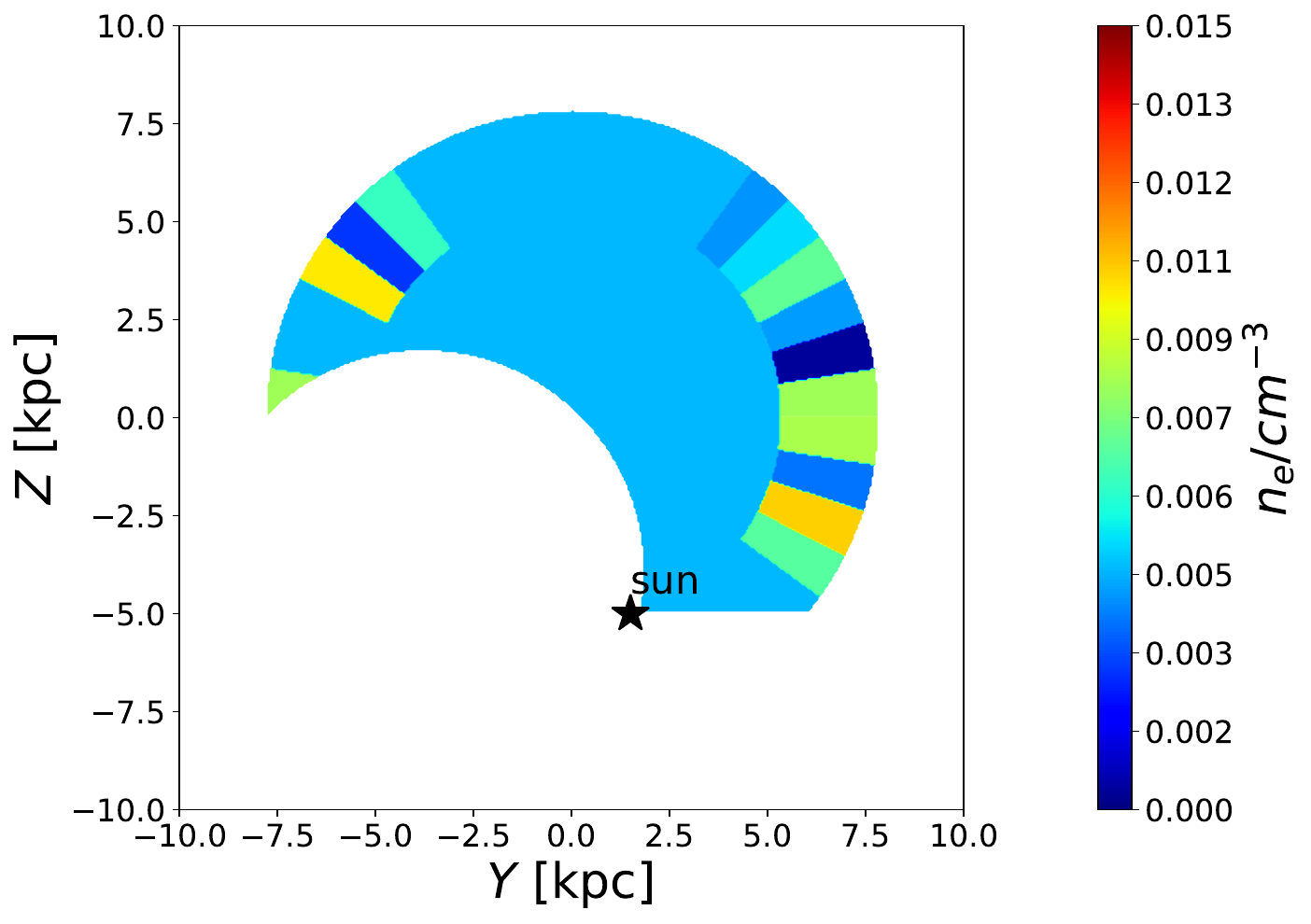}}
	   {\includegraphics[width=0.45\textwidth]{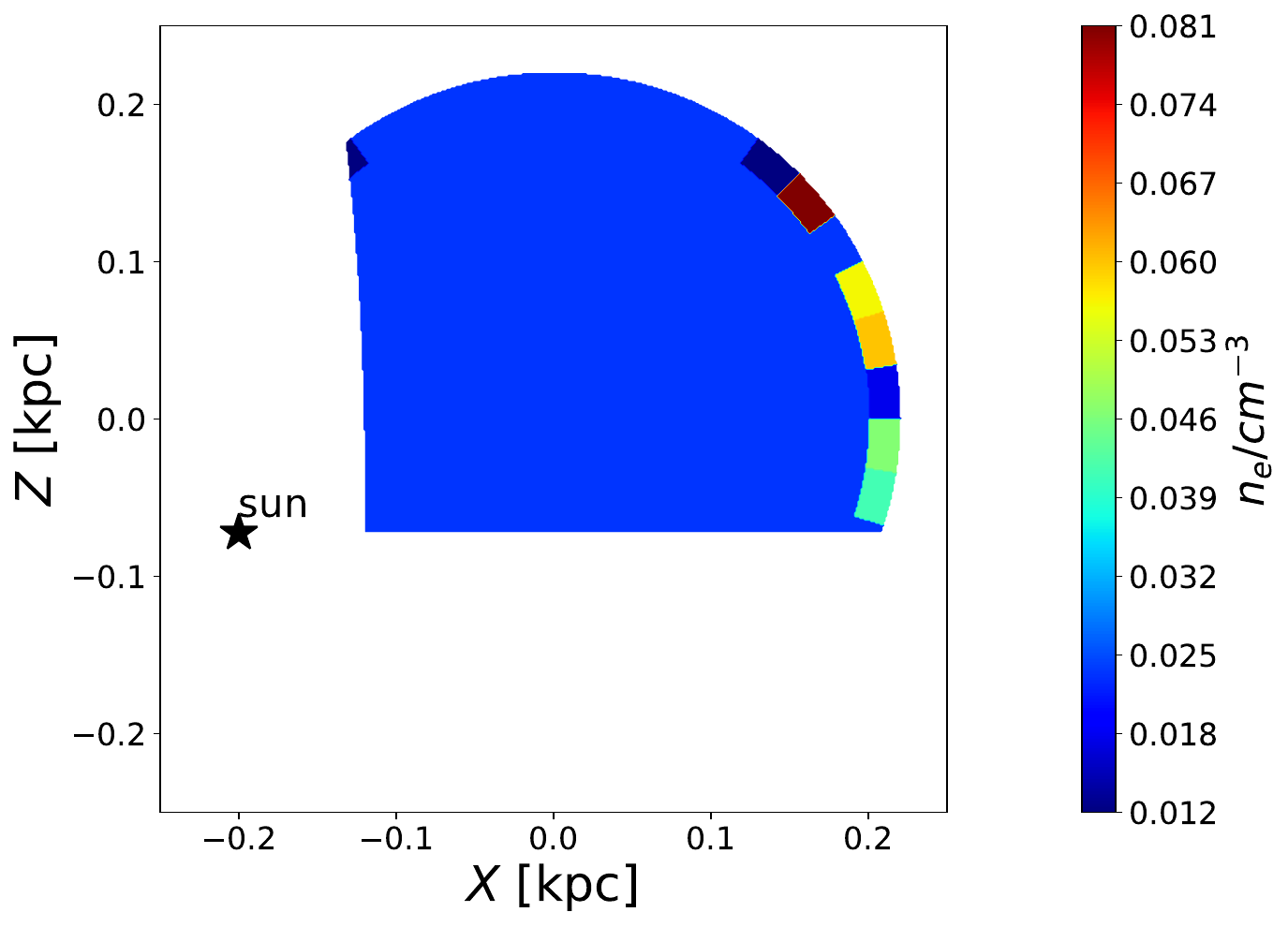}}
	   {\includegraphics[width=0.45\textwidth]{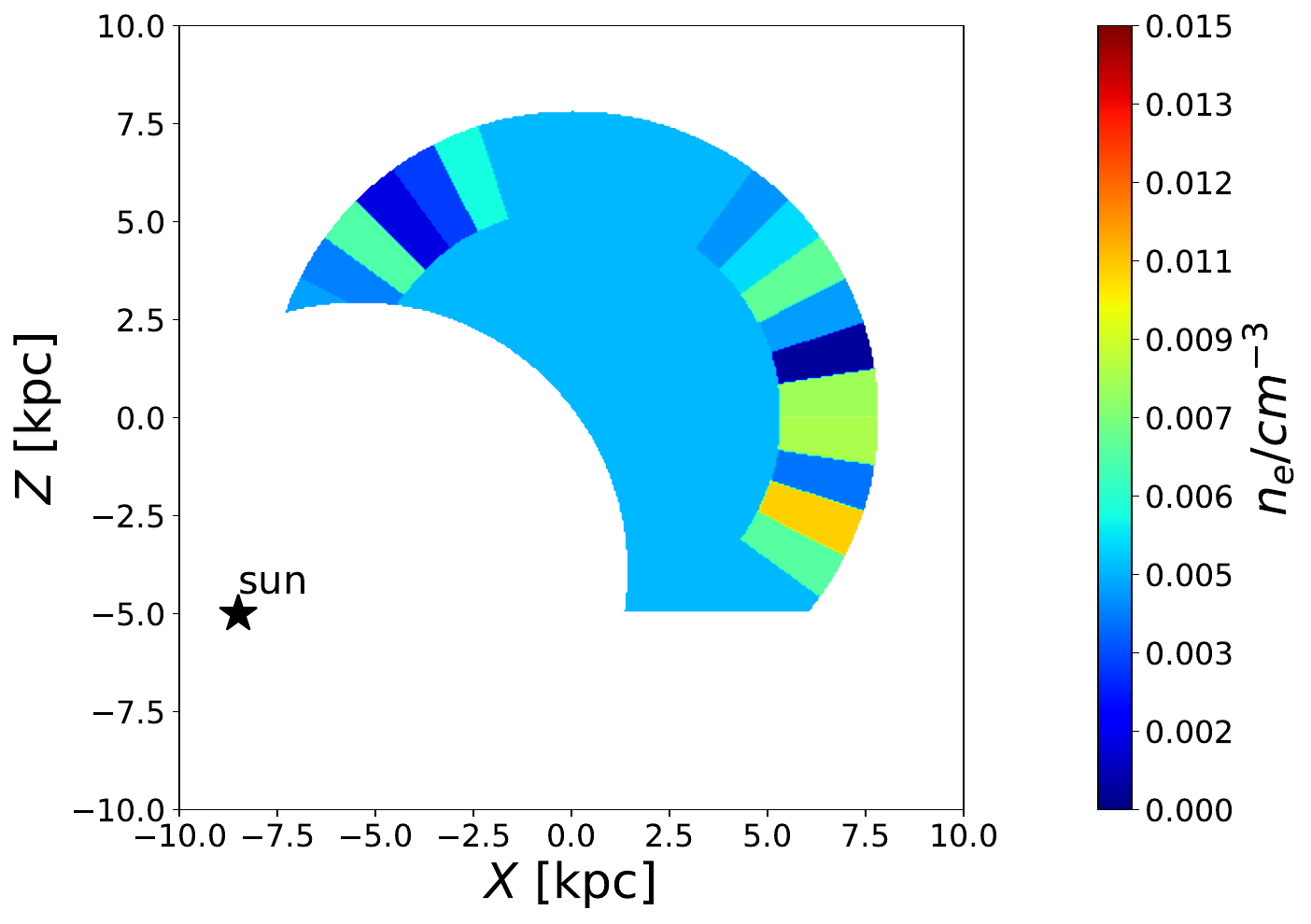}}
	  {\includegraphics[width=0.45\textwidth]{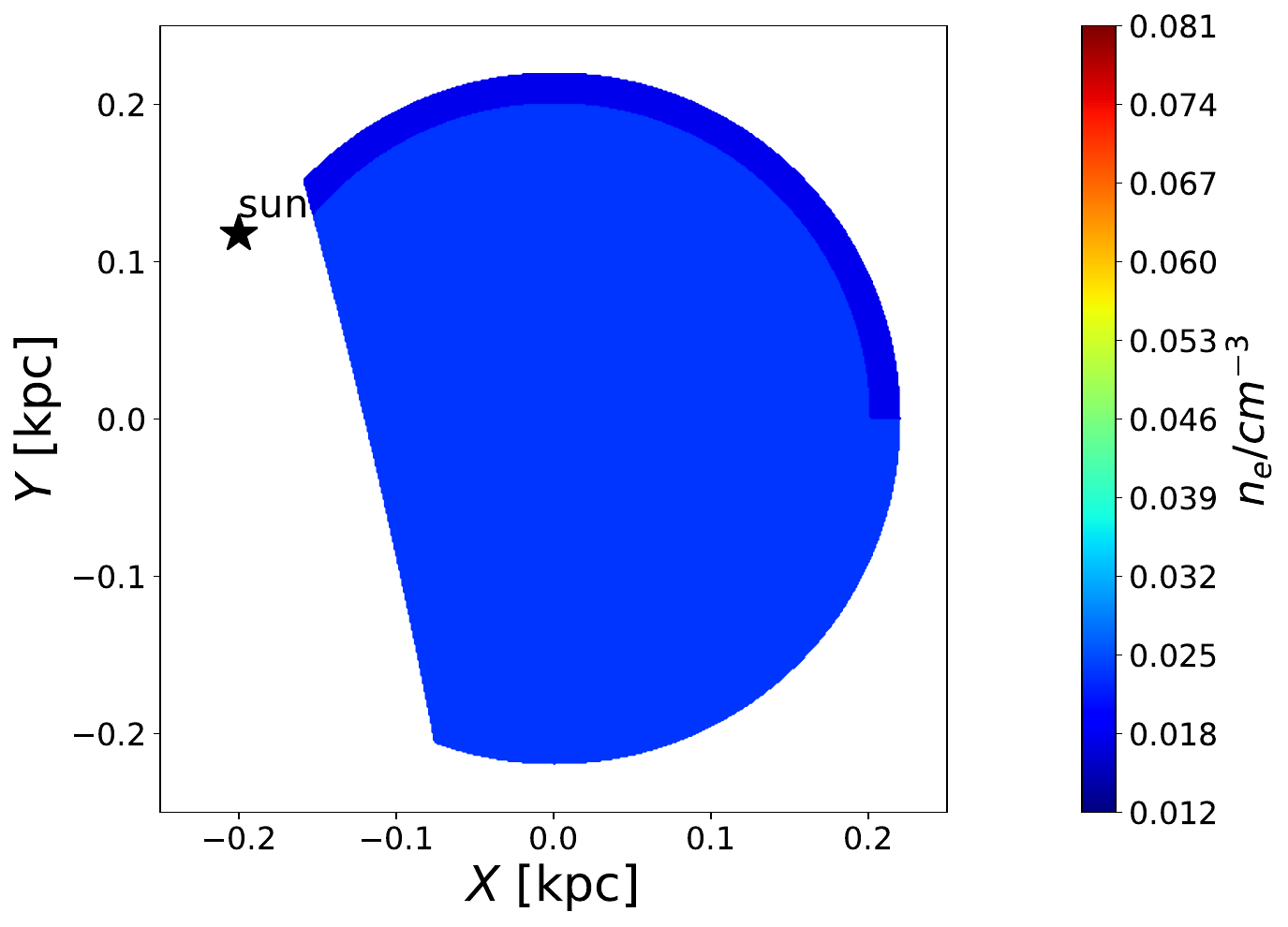}}
	   {\includegraphics[width=0.45\textwidth]{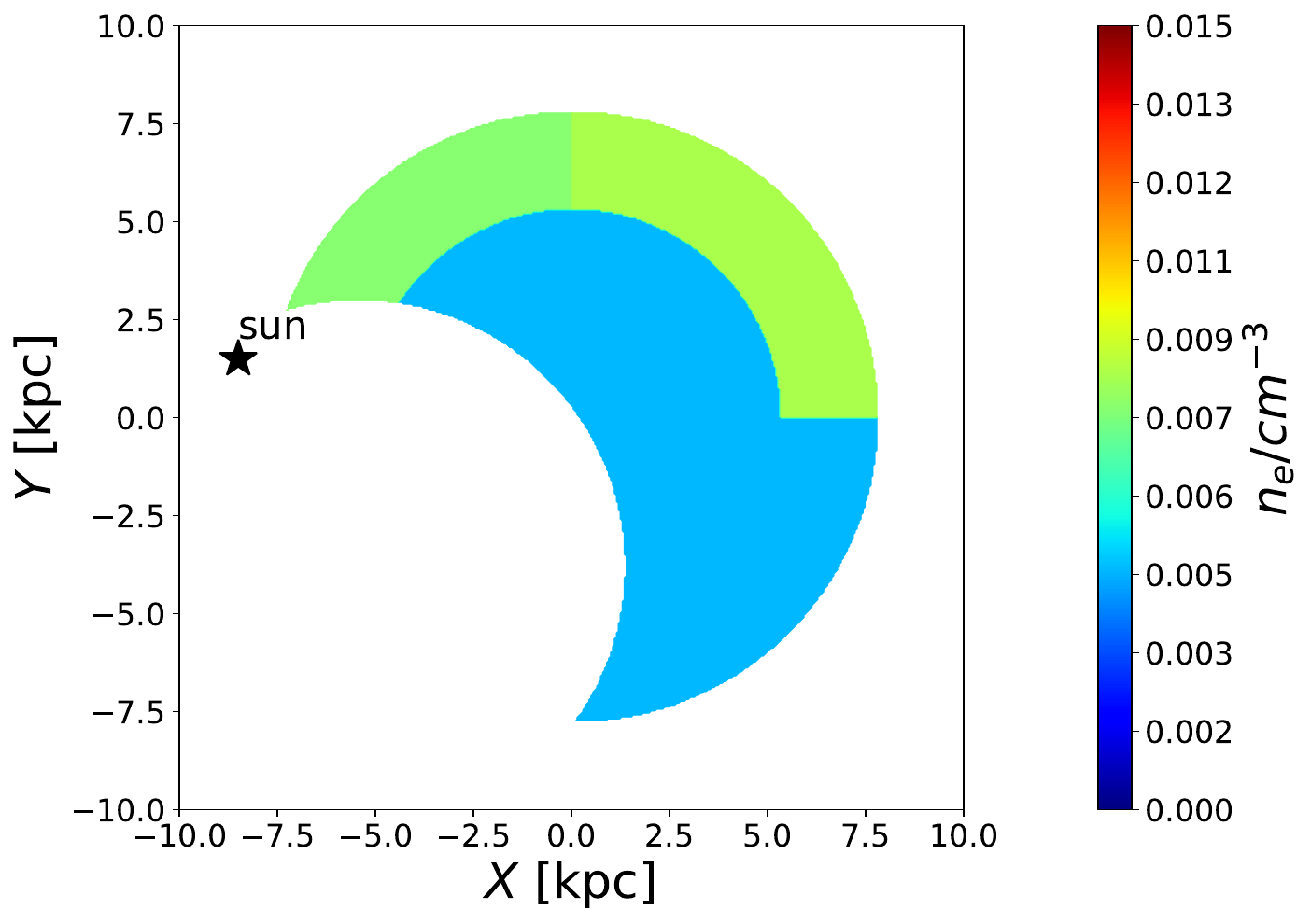}}
	\caption{The Loop I/NPS electron density constructed from {\tt HaloSat} observations for the shell-like SNRs (left) and GC model (right), respectively. From top to bottom, we plot the $X'=0$, $Y'=0$, and $Z'=0$ planes, respectively, in the Loop I/NPS-centered coordinate frame and mark the Sun's position of the projection in each panel. In the left panels, the Sun's position is at ($-0.196, 0.118, -0.072$) kpc, while in the right panels, it is at ($-8.5, 1.5, -5$) kpc.  
	}
	\label{fig:Halosat_ne_slice_in_xyz}
\end{figure*}

\begin{figure*}[t]
	\centering
	 {\includegraphics[width=0.45\textwidth]{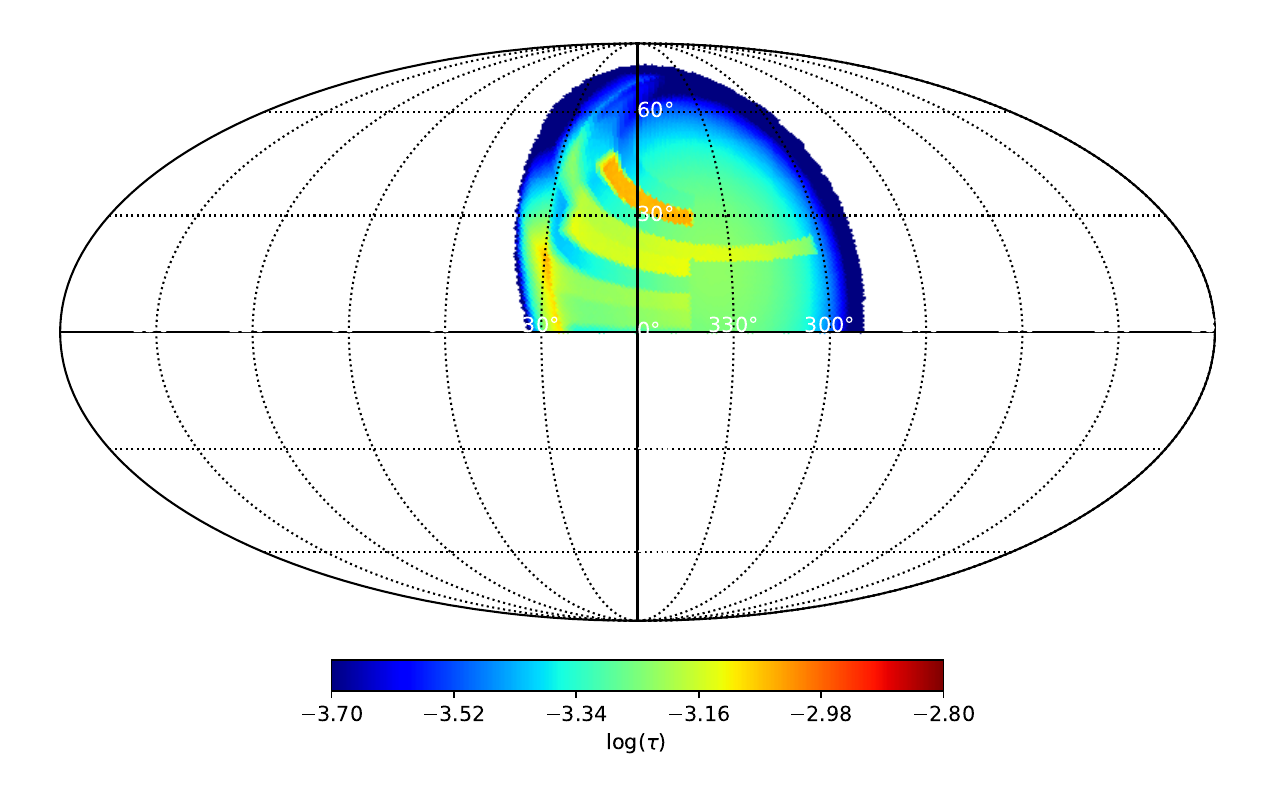}}
	 {\includegraphics[width=0.45\textwidth]{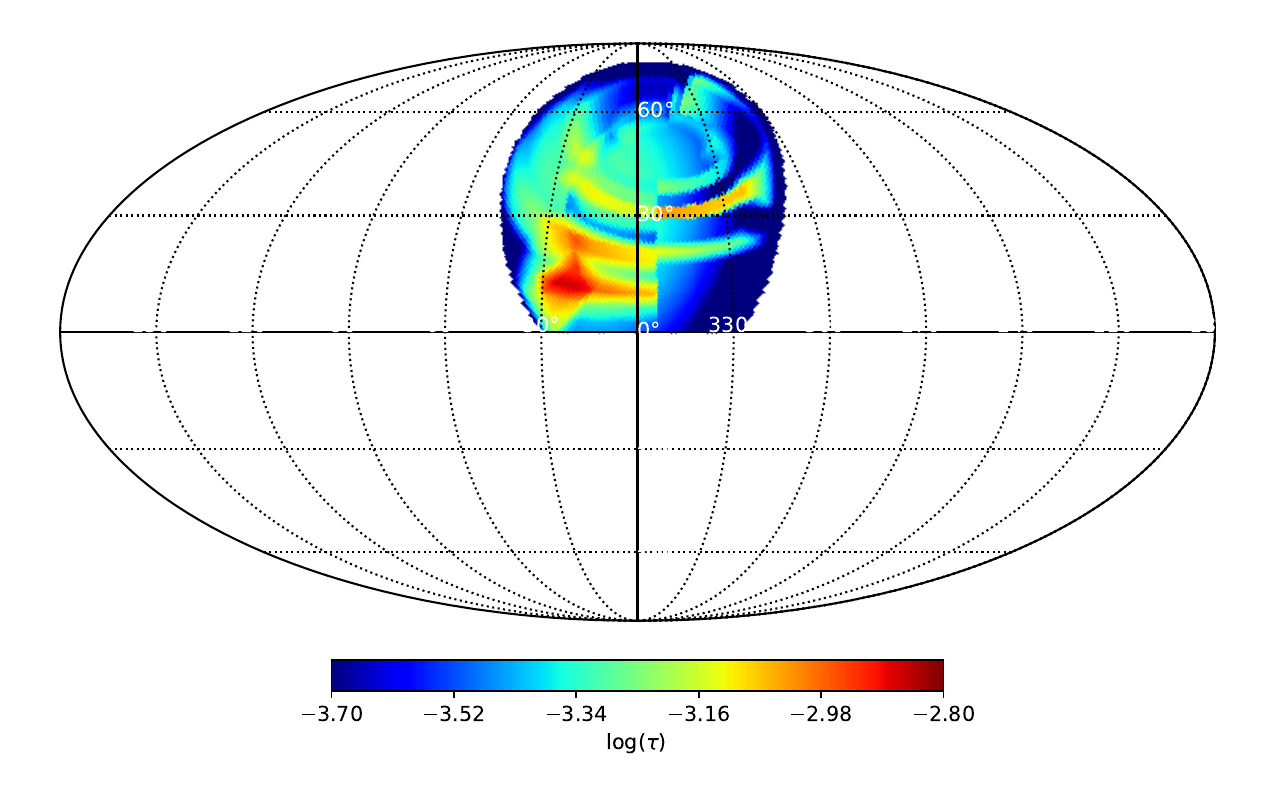}}
     {\includegraphics[width=0.45\textwidth]{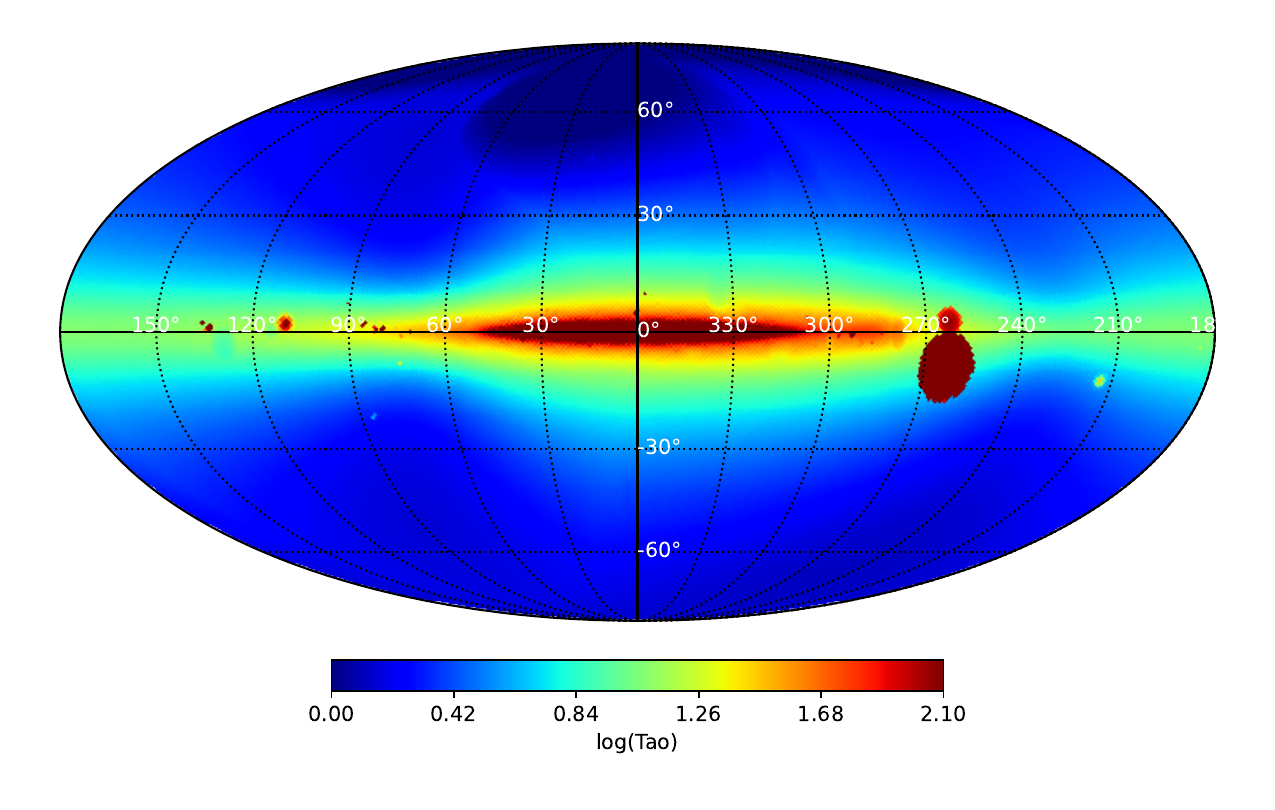}}
     {\includegraphics[width=0.45\textwidth]{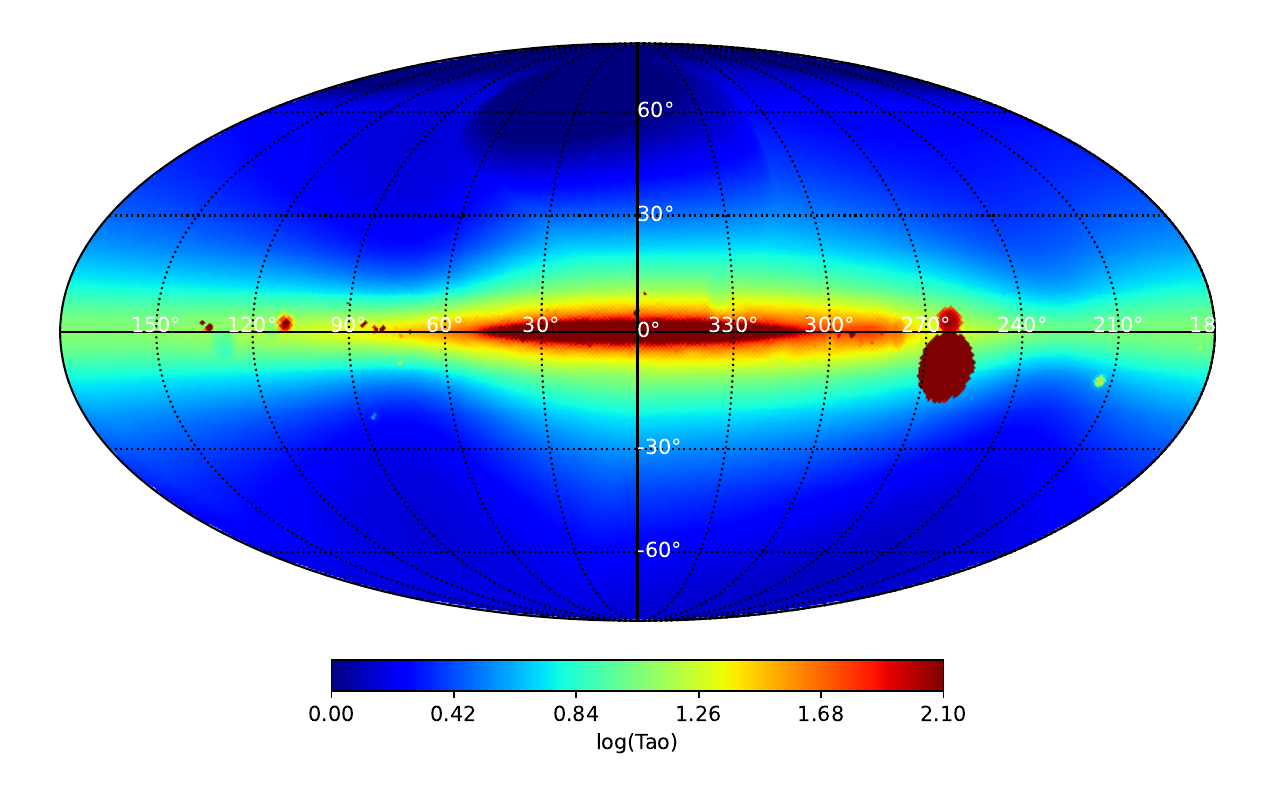}}
	\caption{
    {\black The free-free absorption optical depth  at 1 MHz.
    {\it Left}: the optical depth of the Loop I/NPS in the shell-like SNRs model. {\it Right}: the optical depth in the GC model.
    The top panels solely show the optical depth attributed to the Loop I/NPS itself, while the bottom panels present the total absorption depth integrated over each line-of-sight. 
    }
}
	\label{fig:Tao_at_1MHz}
\end{figure*}

Fig. \ref{fig:Halosat_ne_slice_in_xyz} shows the constructed electron density distribution of the Loop I/NPS for the slices of $X'=0$, $Y'=0$, and  $Z'=0$ in the Loop I/I/NPS-centered coordinate frame.   
{\black The EM is converted into optical depth via \citep{Condon2016erabook}}
\begin{equation}
    \tau_{\nu} \approx 3.28 \times 10^{-7} \left(\frac{T_e}{10^4\thinspace \rm K}\right)^{-1.35}\left(\frac{\nu}{\rm GHz}\right)^{-2.1}\Big(\frac{\text{EM}}{\rm pc \thinspace cm^{-6}}\Big),
\end{equation}
assuming electron temperature $T_e=10^6$ K, for the two Loop I/NPS models.
{\black The optical depth for free-free absorption of the Loop I/NPS solely at 1 MHz is presented in Fig. \ref{fig:Tao_at_1MHz}.}
Interestingly, the optical depth maps do not show loop-like {\black features}, which is consistent with the EM map derived from Planck observations \citep{Plank2020I}.
{\black As a comparison, in Fig. \ref{fig:Tao_at_1MHz} we also show the integrated free-free optical depth for the full Milky Way, including the contribution of the Loop I/NPS and other Galactic electrons. Compared with this total optical depth, the contribution from the Loop I/NPS is negligible and invisible on the maps.}

We find that, the optical depth from the Loop I/NPS itself is small, primarily because the electrons therein are hot.
The EM derived from {\tt HaloSat} observations is for electrons with temperature $\sim10^6$ K, much hotter than the typical WIM  (several thousand K) in the Milky Way. However, it seems that in the Loop I/NPS, electrons with temperature close to the typical WIM are few because in the H$\alpha$ sky map, there is no corresponding loop-like structure found \citep{Haffner2003ApJS}. We check that, if we replace the Loop I/NPS model with a hemisphere with a radius of 7.8 kpc (0.22 kpc), the thickness of the shell 2.5 kpc (0.02 kpc), density of the shell  0.01 cm$^{-3}$ (0.081 cm$^{-3}$) and density of the sphere interior 0.005 cm$^{-3}$ (0.024 cm$^{-3}$), and temperature $10^6$ K. The optical depth is still always lower than $\sim 3\times 10^{-3}$. If the temperature is set to be 8000 K, then the optical depth can reach up to $\sim 2$, reducing the radiation behind Loop I/NPS to $\sim 14\%$.   
In summary, the self-absorption of the Loop I/NPS emission should have modest effects on the final sky map.

\section{Results and Discussion}\label{sec:result}
 
\subsection{Results}

\begin{figure*}[t]
	\centering
	 {\includegraphics[width=0.45\textwidth]{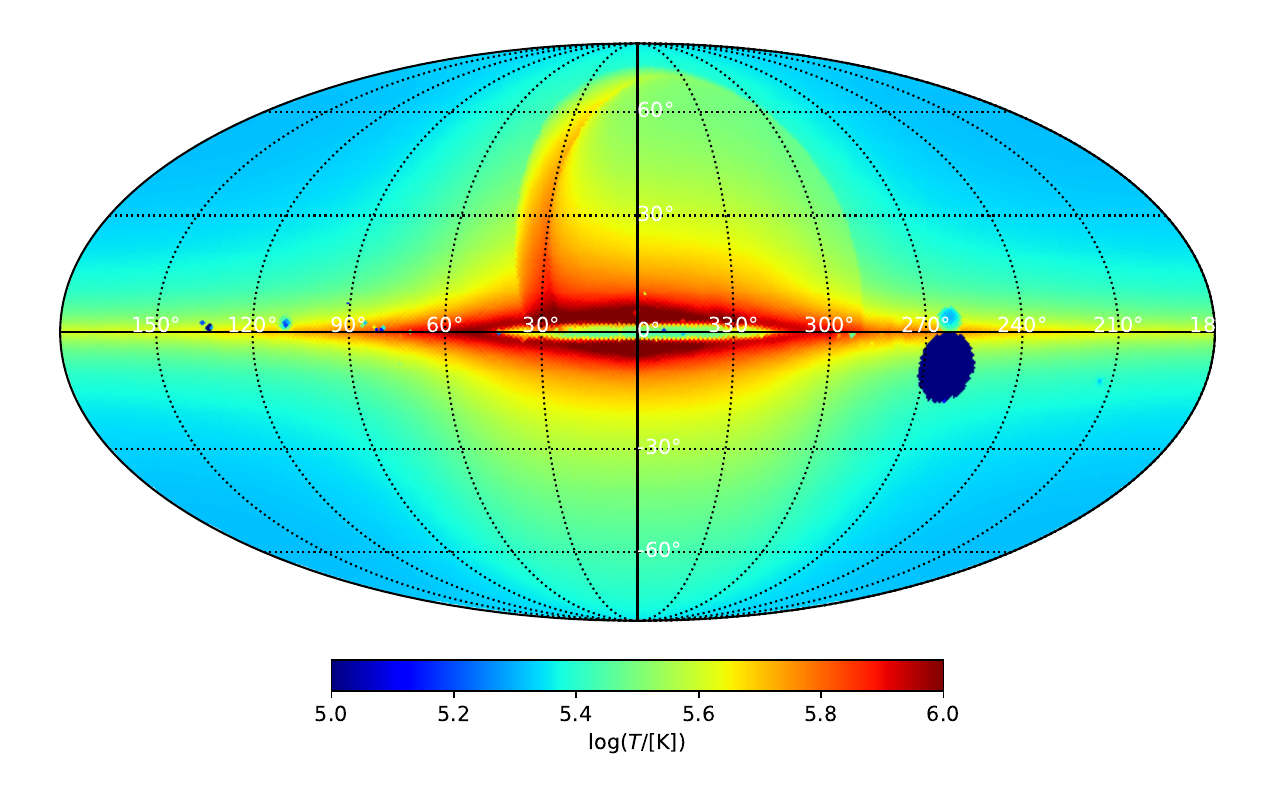}}
     {\includegraphics[width=0.45\textwidth]{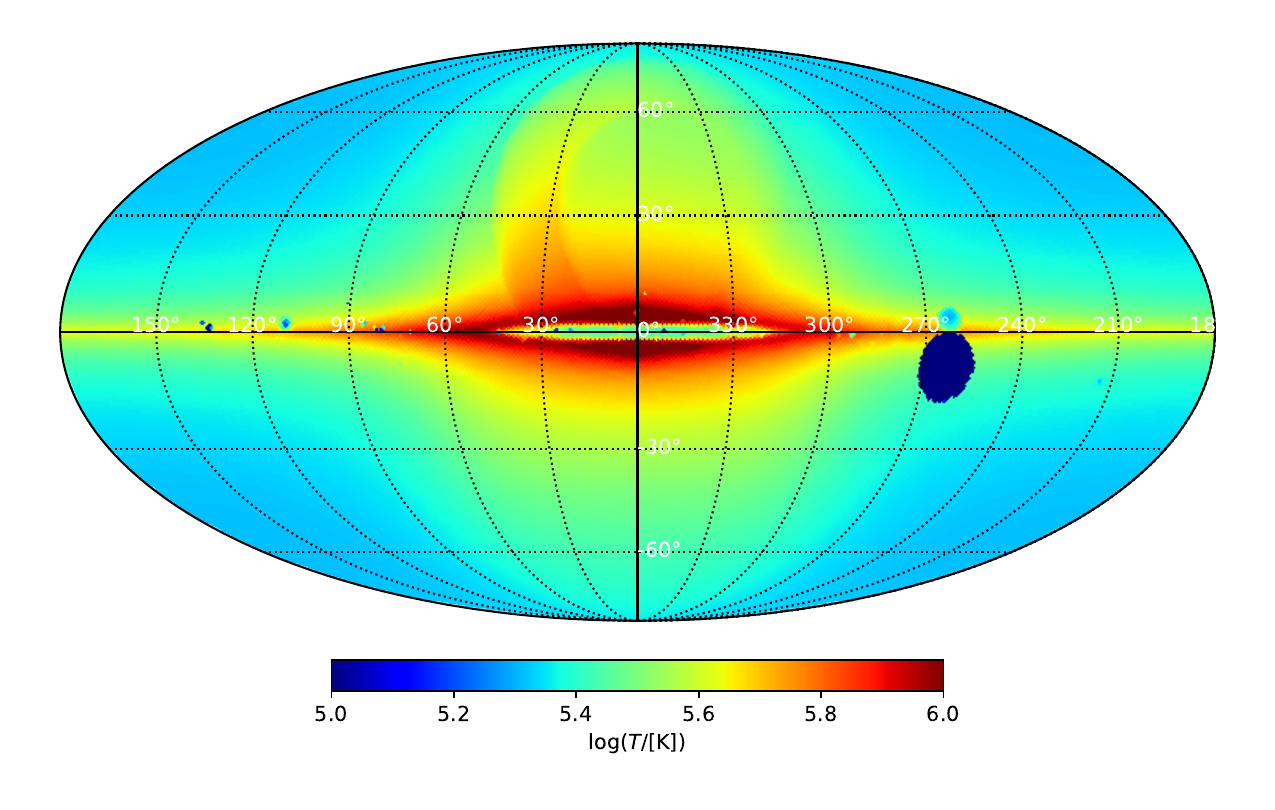}}
	 {\includegraphics[width=0.45\textwidth]{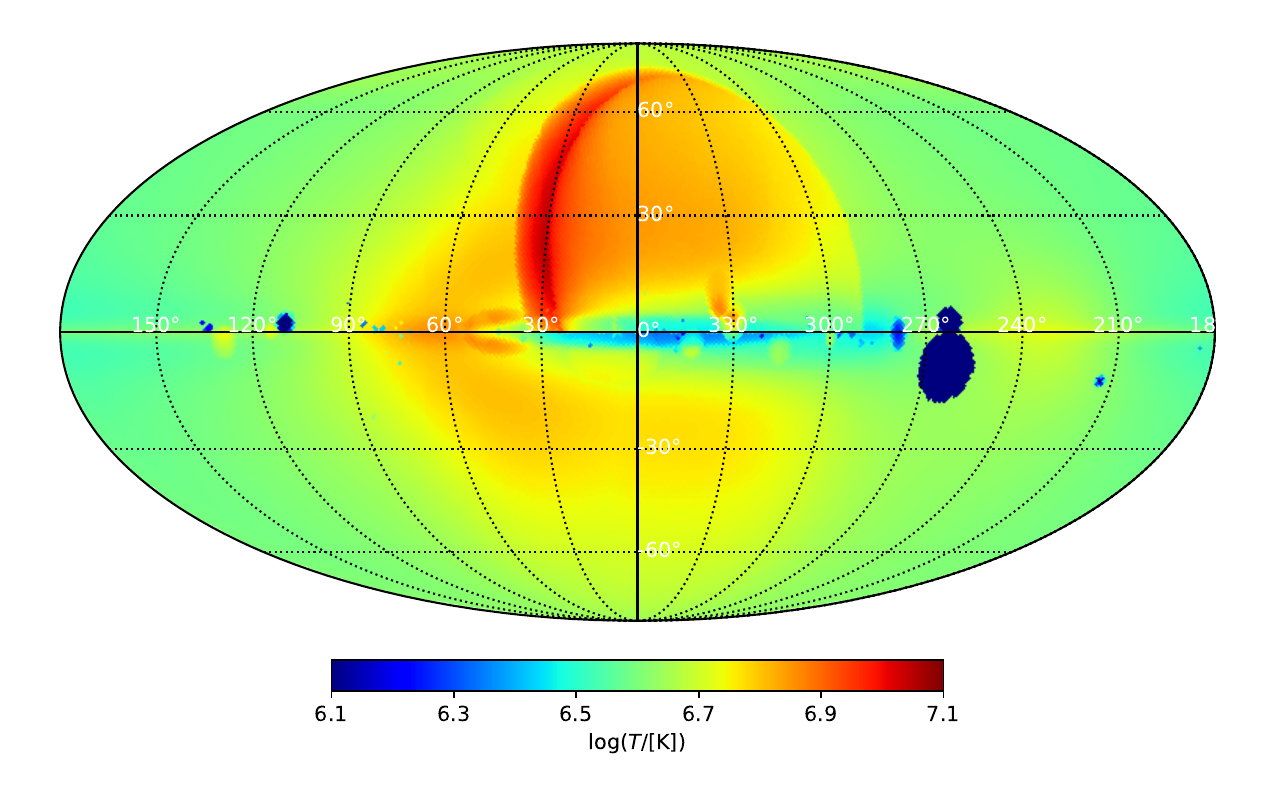}}
     {\includegraphics[width=0.45\textwidth]{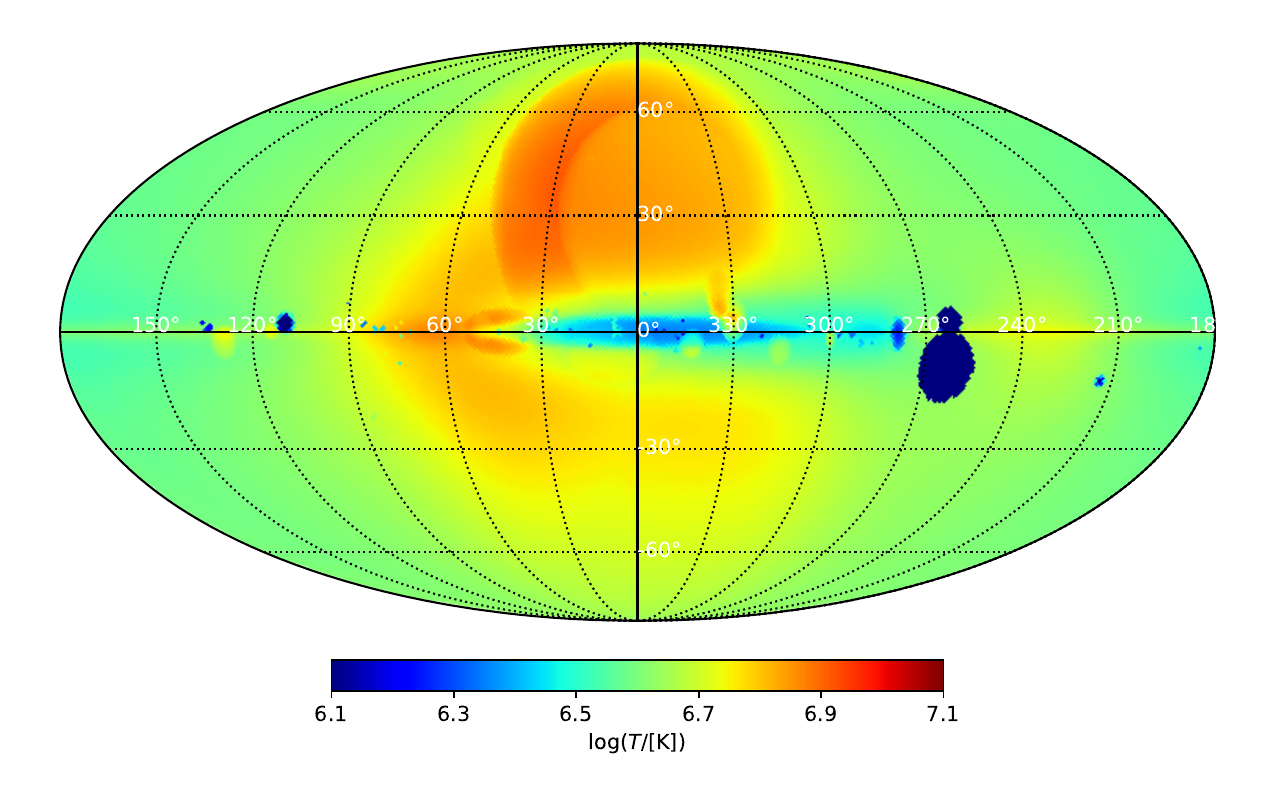}}
	 {\includegraphics[width=0.45\textwidth]{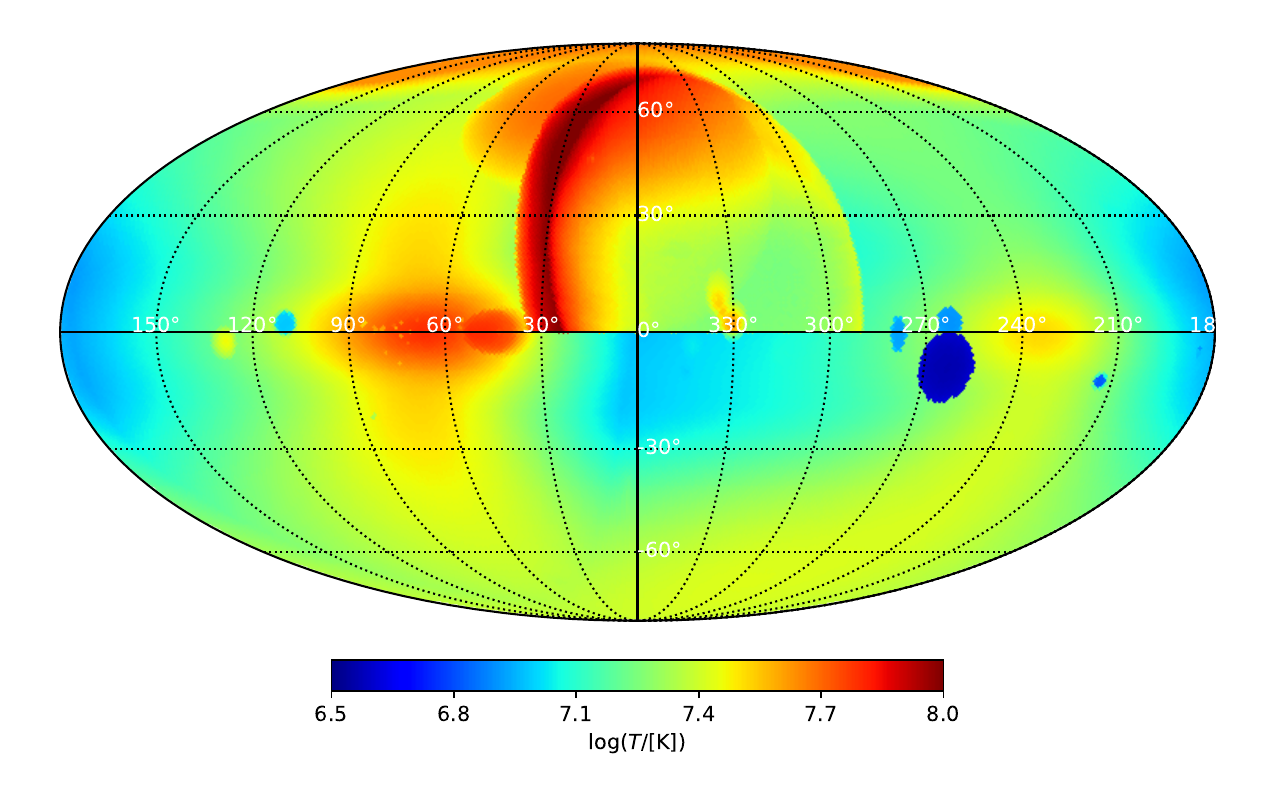}}
     {\includegraphics[width=0.45\textwidth]{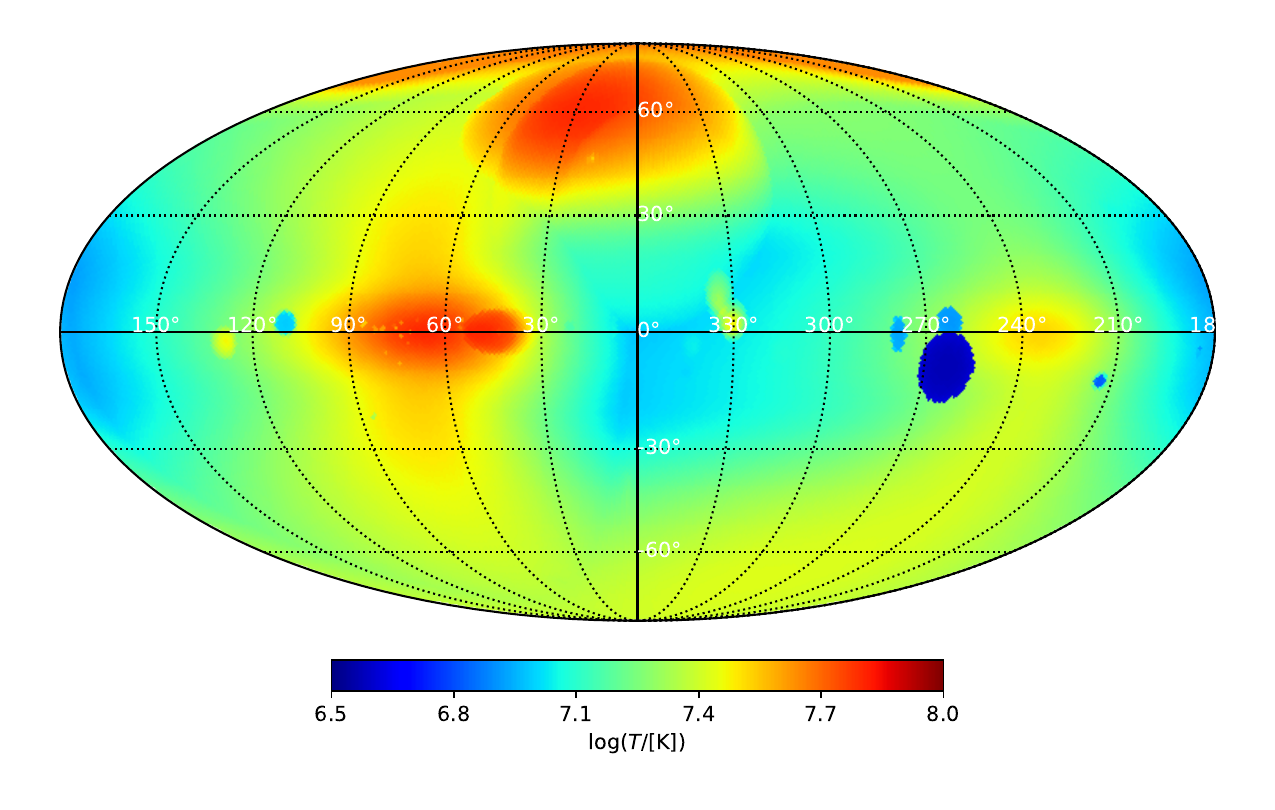}}
        \caption{The predicted sky maps at frequencies 10, 3, and 1 MHz ({\black from top to bottom}) respectively, for the shell-like SNRs model of the Loop I/NPS (left column) and the GC model (right column). 
        {\black We use the {\tt NE2001} model for the free electron distribution.       
        If the Loop I/NPS is a nearby object, it will still be fully visible even at $\sim 1$ MHz. However, if it is as far as the GC, then it vanishes at low latitudes due to absorption.        
        }
}\label{fig:absorption_skymap_use_ne2001}
\end{figure*}

With the modeled 3D emissivity and free electron distribution, the sky map is obtained by performing integration of the radiative transfer function along each {\black line-of-sight}.
Considering the free electron distribution within the Milky Way, radiation from the Loop I/NPS would experience absorption in the ultra-long wavelength band. If the Loop I/NPS is a complex of nearby SNRs, the absorption between Loop I/NPS and the observer would be small; whereas if it is located near the GC, the influence of absorption by the ISM would be significant. Our calculations validate these expectations.
In Fig. \ref{fig:absorption_skymap_use_ne2001} we present the predicted sky maps at frequencies 10, 3, and 1 MHz, respectively, for our two Loop I/NPS origin models. Clearly, below $\sim 10$ MHz, particularly at the frequency as low as 1 MHz,  the high Galactic latitude regions are brighter while the Galactic disk is darker. However, In Fig. \ref{fig:absorption_skymap_use_ne2001}, on the Galactic disk, at the directions where the projection of electron density is low in the {\tt NE2001} model, say the LDR (low-density region, $30\degree \lesssim l \lesssim 90\degree$) and the LSB (local superbubble around $l \sim 240\degree$ ), there are bright patches.
This is already pointed out in \citet{Cong2021APJ}.

At frequencies $\gtrsim 3$ MHz, in both shell-like SNRs and GC models, the Loop I/NPS still appears as a bright arc. {\black But there are differences at frequencies as low as $\sim 1$ MHz. The full Loop I/NPS is still a bright structure in the shell-like SNRs model.} This is because it is very close to the observer, and the absorption by the ISM electrons is negligible. However, in the GC model, the morphology is quite different. The Loop I/NPS disappears at $b \lesssim 30\degree$, due to the absorption of electrons between the GC and the observer. So indeed, the ultra-long wavelength observations can distinguish the models for Loop I/NPS's origin. We also note that, in both the SNRs model and GC model, at $b \gtrsim 30\degree$, the Loop I/NPS is always a bright structure. It means that the absorption toward the high Galactic latitudes, even {\black up to above} the GC, is still weak at least {\black above} 1 MHz. This is further helpful for deriving the electron distribution {\black along the vertical ($Z$) direction, i.e., height of the Galactic disk}. 

\begin{figure*}[t]
	\centering
	 {\includegraphics[width=0.45\textwidth]{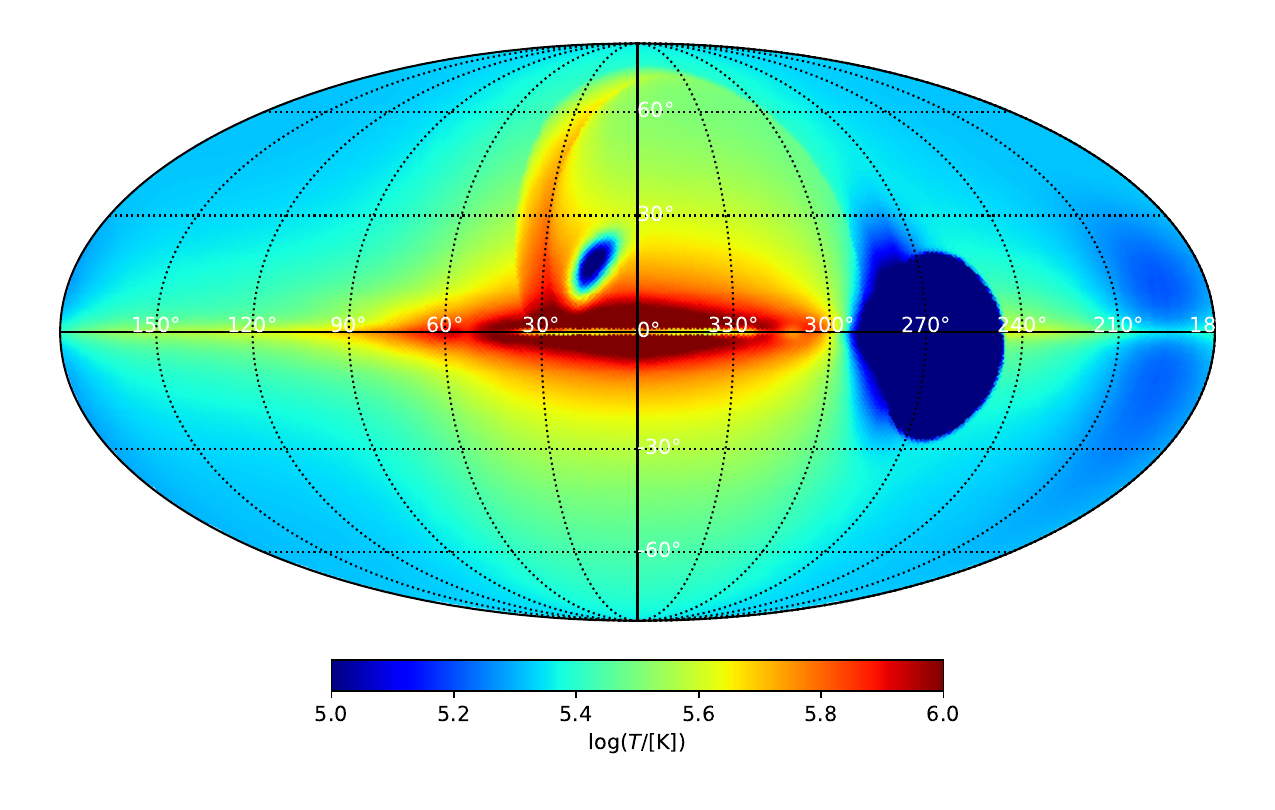}}
     {\includegraphics[width=0.45\textwidth]{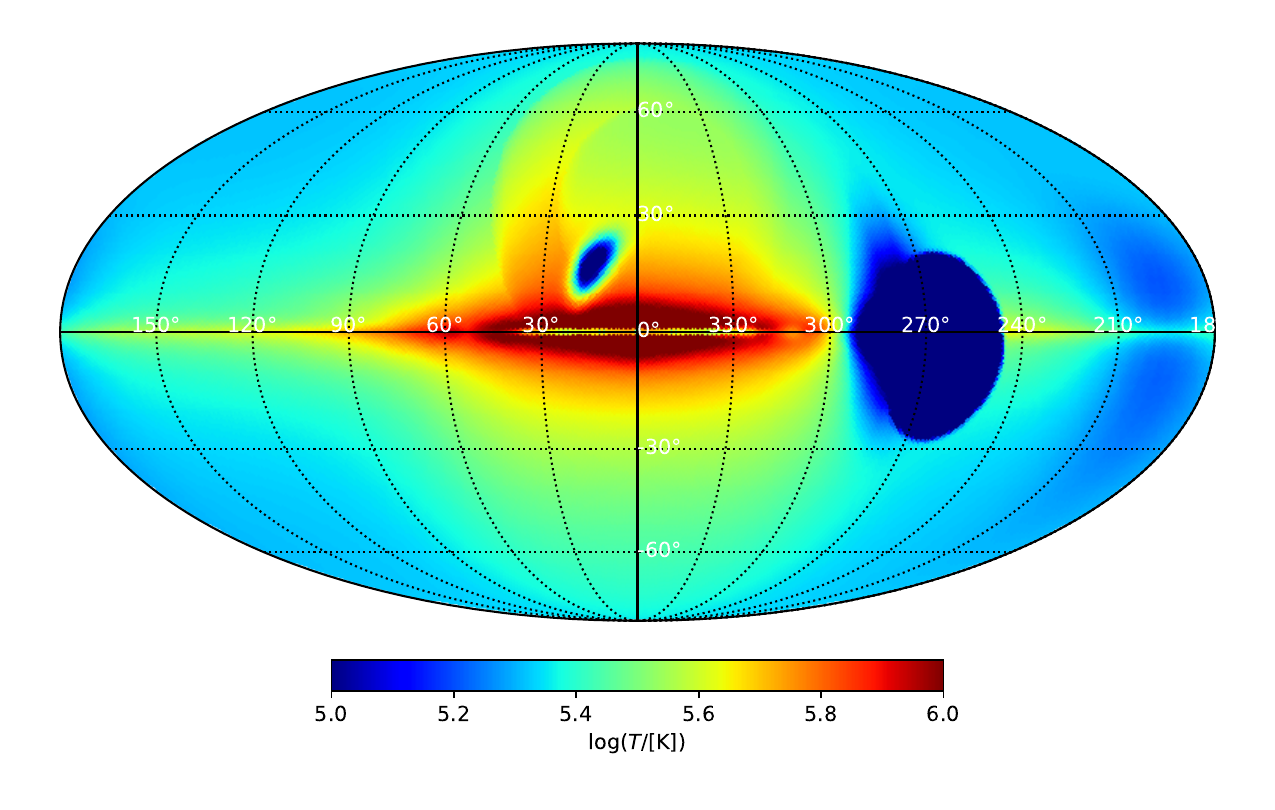}}
	 {\includegraphics[width=0.45\textwidth]{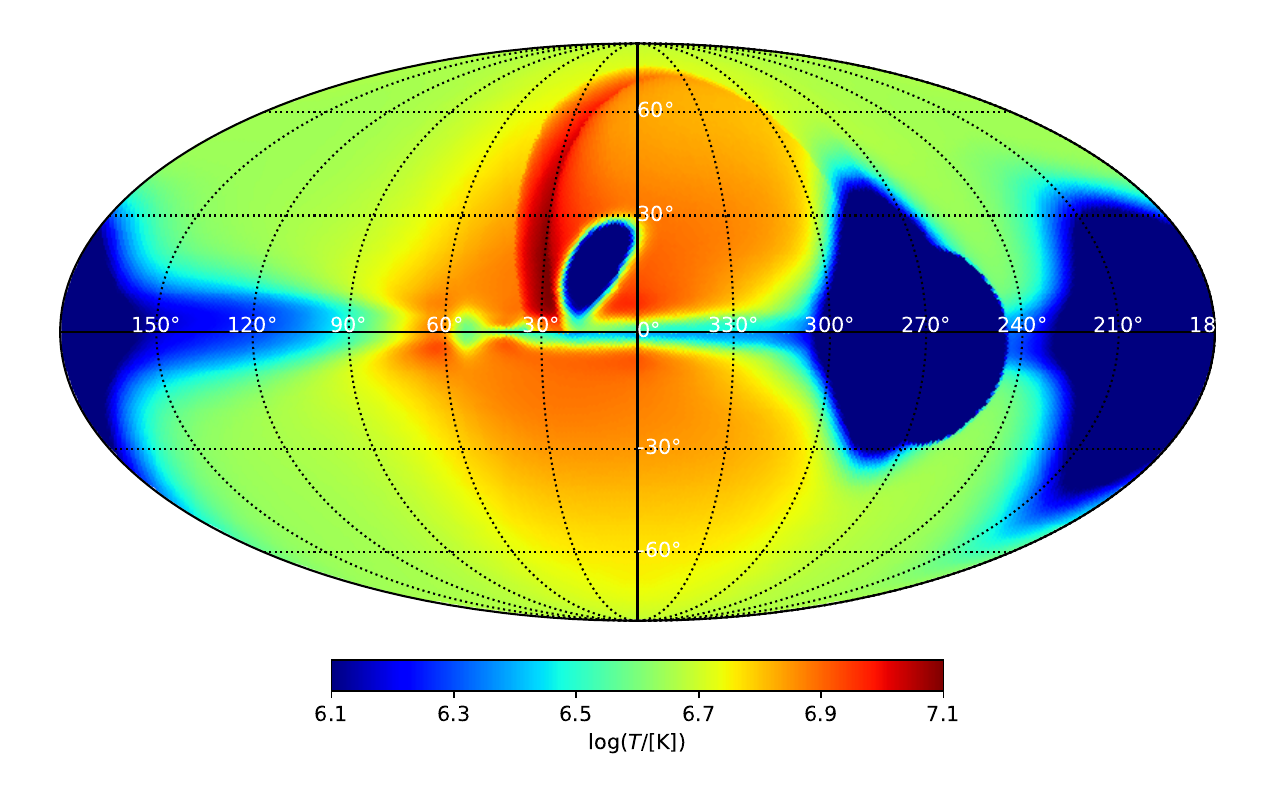}}
     {\includegraphics[width=0.45\textwidth]{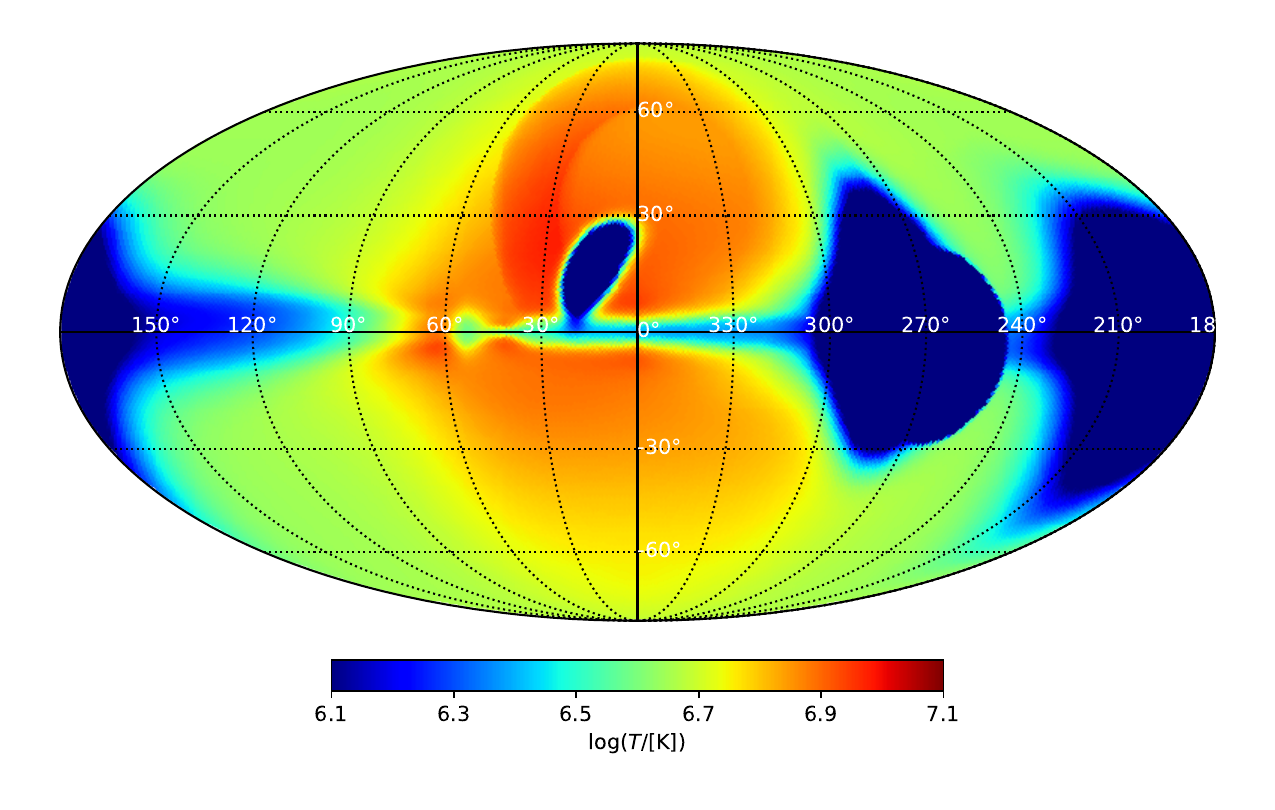}}
	 {\includegraphics[width=0.45\textwidth]{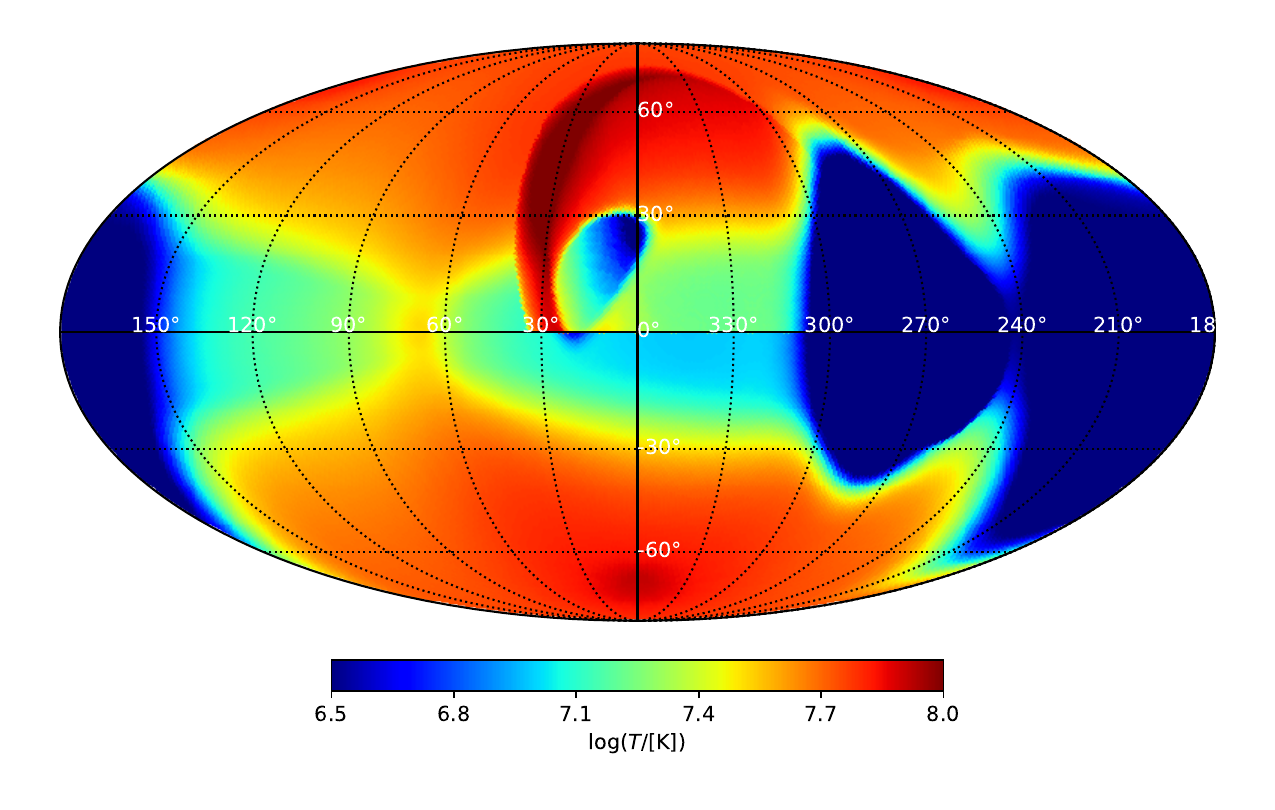}}
     {\includegraphics[width=0.45\textwidth]{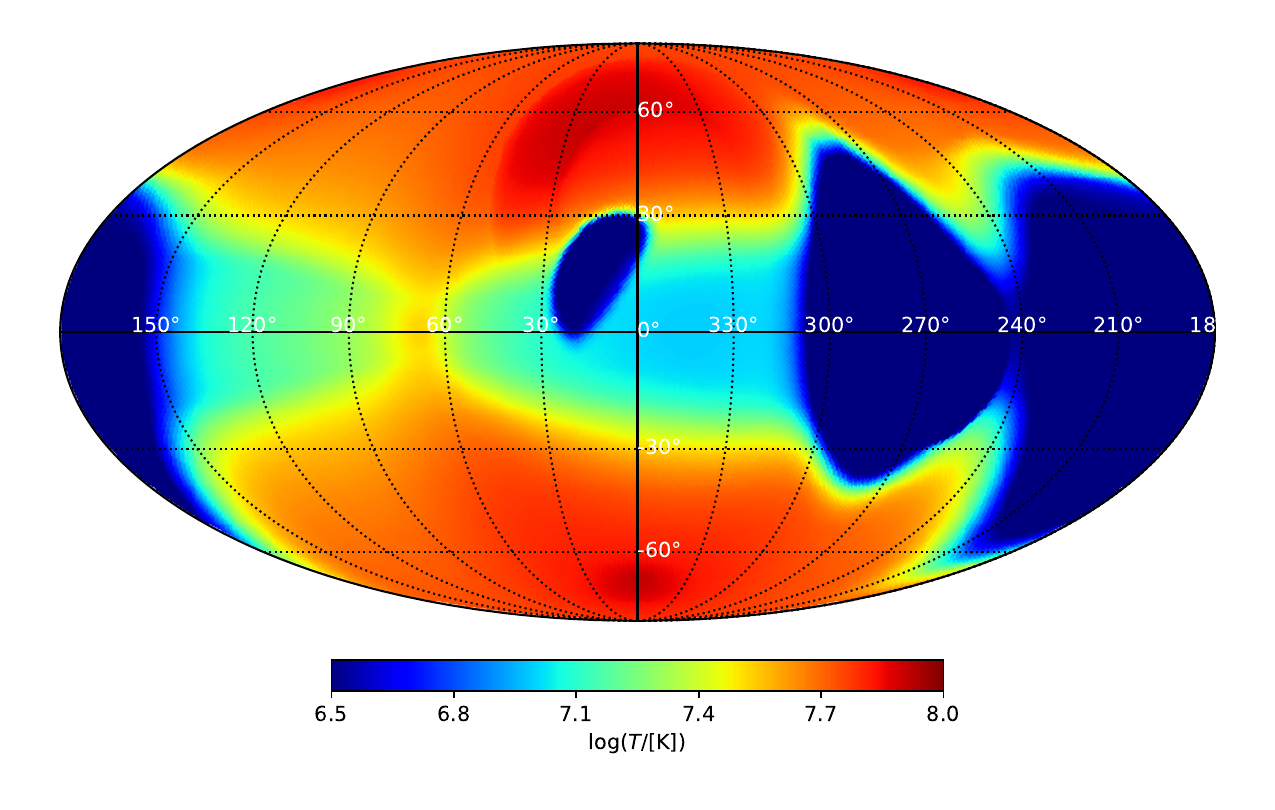}}
        \caption{{\black 
        Similar to Fig. \ref{fig:absorption_skymap_use_ne2001}, however, the {\tt YMW16} model is adopted for the electron density.
        }
        }\label{fig:absorption_skymap_use_ymw16}
\end{figure*}

{\black
Besides the {\tt NE2001}, {\tt YMW16} is a more up-to-date Galactic electron model \citep{Yao2017ApJ}, which uses more recent data to optimize the model. {\tt YMW16} incorporates a number of local dense clumps/voids, e. g. the Gum nebula, the LHB, the Loop I/NPS, etc., but excludes those less important clumps/voids listed in {\tt NE2001}. 
This model exhibits a higher electron density than {\tt NE2001} for the Galactic plane, 
Loop I and Gum nebula regions.
Both models utilize a spherical shell to reconstruct the morphology of the Loop I/NPS. Our predictions for the ultra-long wavelength sky using the {\tt YMW16} electron density model are shown in Fig. \ref{fig:absorption_skymap_use_ymw16}. We have calibrated the fluctuation parameter so that  the predicted global spectrum of the radio background for {\tt YMW16} is similar to that for {\tt NE2001}, both consistent with observations. For {\tt YMW16}, the calibrated fluctuation parameter is 0.3, while for {\tt NE2001}, we basically use 3.0. The results for {\tt YMW16} are generally similar to {\tt NE2001}. However, there are some differences: first, the absorption is much stronger around the Galactic plane in {\tt YMW16}. Even at $\sim 3$ MHz, a large fraction of the regions around the Galactic plane becomes darker. 
Second, with {\tt YMW16} at 1 MHz, even for the SNRs model the root of the Loop I/NPS is absorbed a bit. This is because in {\tt YMW16} the Loop I/NPS itself is thick and dense, with the highest density $\sim 3~{\rm cm}^{-3}$. Third, at the Galactic plane near $l\sim 60^\circ$, there is a big and bright spot in {\tt NE2001}, however, in {\tt YMW16} there is no such feature. This is because {\tt NE2001} has a giant low-density bubble in this direction, therefore the absorption is much weaker than the ambient medium. However, the primary conclusion based on both of the two models is the same: if the Loop I/NPS is a nearby object, it will be fully visible at $\sim 1$ MHz. However, if it is close to the GC, then only the high Galactic latitude parts at $b\gtrsim 30^\circ$ would be still visible at such low frequencies.
}

\subsection{Discussion}

 We did not employ any physical hypothesis on the origin of Loop I/NPS, which is still an open question. We just adopted phenomenological models to describe its location, size, and emissivity. Our forecast on the Loop I/NPS morphology in the ultra-long wavelength band should be robust. However, we note that at different wavelengths the Loop I/NPS morphology is different. In the radio band, it is an arc above the Galactic plane; however, in the soft X-ray band it extends to below the Galactic plane and forms dual bubbles \citep{Predehl2020Natur}. 
 In the polarization map, there are many spurs inside the Loop I/NPS (above and below the Galactic plane), and they are all rooted on the Galactic plane \citep{Bennett2013ApJS,Planck2016XXV,Dickinson2018Galax}. 
 Since our phenomenological model only aims to reproduce the observed Loop I/NPS morphology at 408 MHz, it is possible that in the ultra-long wavelength band, there are new structures near the Loop I/NPS. They may confuse our identification of the Loop I/NPS origin. 

Regarding the Loop I/NPS, there are arguments supporting both a local feature and a GC feature. It is also possible that a local feature coincides with a GC feature.
If the Loop I/NPS comprises two visually overlapping components, three hypotheses can be considered in this context. The first one is that the high Galactic latitude portion of Loop I/NPS is located at GC, while the remaining segment originates from a nearby source. Due to weaker absorption at high Galactic latitudes, the Loop I/NPS would still manifest as a bright, complete arc in the ultra-long wavelength band, rendering it indistinguishable from the shell-like SNRs model.
The second hypothesis proposes that the high Galactic latitude portion is a nearby structure, while the remaining segment resides at the GC. Due to stronger absorption at low Galactic latitudes, only the high Galactic latitude regions are visible ultra-long wavelength band, making it indistinguishable from the GC model.
The final hypothesis is an overlap of nearby and GC components throughout the Loop I/NPS. If the entire loop is observable at ultra-long wavelengths, the high-latitude section would consist of both nearby and GC components, while the low-latitude region would predominantly originate from nearby emission. This scenario may remain indistinguishable from the shell-like SNRs model.

{\black 
In the 1.4 GHz polarization intensity map \citep{Reich2009IAUS}, apparently, the low-latitude part ($b\lesssim 30-40\degree$) is entirely different from higher latitudes because the  Faraday depth and Faraday rotation depolarization amount become much larger \citep{Sun2015ApJ,Dickinson2018Galax} at low latitudes. However, at higher frequencies (several tens of GHz), the absorption and Faraday depolarization effects are negligible even near the Galactic plane. Therefore, polarization is a more useful tool for assessing whether the high-latitude part and low-latitude part of the Loop I/NPS belong to the same structure.  As if the Loop I/NPS is composed of two physically unrelated and spatially distinct components, it will be unlikely to observe a coherent polarization structure. \citet{Vidal2015MNRAS} and \citet{Planck2016XXV} analyzed the WMAP and Planck observations and found that in sky maps of both polarization and projected magnetic field, the entire Loop I/NPS is a coherent structure. \citet{Planck2016XXV} also found that this polarization structure extends beyond the Loop I/NPS boundary. 
Although in a closer inspection of their projected magnetic field angle sky map (lower panel of their Fig. 20), it seems that the part close to the Galactic plane is separated from the higher latitudes part, these two parts are still difficult to disentangle.
}
 
{\black 
The synchrotron spectral index could depend on the frequency (e.g. \citealt{Huang2019SCPMA,Cong2021APJ,Padovani2021A&A,Irfan2022MNRAS}). 
In \citet{Huang2019SCPMA}, they find the spectral index tends to become flat below $\sim20$ MHz. Despite of this, in this paper, we adopt a constant spectral index for synchrotron emissivity because: 1) Currently the observations below $\sim20$ MHz suffer from large uncertainties,  it is a risk of extrapolating the frequency-dependence of spectral index constrained mainly by higher frequency observations to below $\sim 20$ MHz, and down to as low as $\sim 1$ MHz. 2) In the ultra-long wavelength band, the morphology of the Loop I/NPS is mainly determined by the free-free absorption (i.e., distribution of free electrons, location and geometry of the source), rather than the spectral index of synchrotron emissivity before absorption. Nevertheless, we check that, using a shallower/steeper value of $\beta_{\rm G}$, or even using the frequency-dependent form 
 (Eq. (28) of \citealt{Cong2021APJ}), for synchrotron spectral index, our conclusion will not change.  
}

{\black
The synchrotron spectral index also depends on the direction (e.g. \citealt{Guzman2011A&A,Cong2021APJ}).  In \citet{Guzman2011A&A} it seems that the spectra in $-130\degree \lesssim l \lesssim 90\degree$ are steeper, both above and below the Galactic plane. 
}
In our paper, the spectral indices of $\epsilon_{\rm disk}$ and $\epsilon_{\rm L}$ are the same. Therefore, the change of morphology other than the simple extrapolation from the power-law spectrum is attributed to the free-free absorption. However, if in the Loop I/NPS region, the dependence of spectral indices on Galactic latitudes is different from other regions, i.e., sharper at high Galactic latitudes and shallower at low Galactic latitudes, it may also result in morphology analogous to free-free absorption. However, we check that, at least {\black above} $\sim 20 $ MHz, there is no such {\black phenomenon}, see also  \citet{Guzman2011A&A}. 
{\black
Moreover, both in \citet{Guzman2011A&A} and \citet{Cong2021APJ}, the spectra in the Loop I/NPS region, not just on the bright arc but also inside the loop, are slightly steeper. If this is true, then in the ultra-long wavelength band, the Loop I/NPS will be brighter in case the absorption is negligible. Comparatively, the absorption effects will be more conspicuous. This will strengthen our conclusion.
}

\section{Summary}\label{sec:summary}

In this paper, we investigated the Loop I/NPS structure feature in the ultra-long wavelength band. We considered two models: the shell-like SNRs model, in which the Loop I/NPS is a complex of SNRs close to our Sun, and the GC model, in which it is located close to the Galactic Center and has a large physical size. Both models can reproduce the observational feature (the loop-like feature) at high frequencies where the free-free absorption is negligible. {\black However, the loop-like feature differs significantly in the two models in the ultra-long wavelength band where the free-free absorption is efficient ($\nu \lesssim 3$ MHz).} In the shell-like SNRs model, the Loop I/NPS is still bright at frequencies as low as 1 MHz. However, in the GC model, due to the free-free absorption by the ISM between the GC and the Sun, the Loop I/NPS structure becomes dark or even fully disappears at $b \lesssim 30 \degree$. However, at high Galactic latitudes, $b \gtrsim 30\degree$, the Loop I/NPS structure is still a notable feature in the sky. The upcoming ultra-long wavelength observations, with for example the DSL \citep{Chen2021RSPTA,Chen2024arXiv}, FARSIDE \citep{Burns2020AAS}, etc.,
have the potential to distinguish these two models.

{\black The Loop I/NPS model developed in this paper will be incorporated in the Version 2.0 of our sky model, i.e. ULSA-v2.0. We plan to make it publicly available as long as the new version is ready. Before that, the code can be obtained by making a reasonable request to the authors.
}

\section*{Acknowledgments}
{\black We thank the anonymous referee very much for giving us detailed suggestions that greatly improve the manuscript.}
This work was supported by the National Key R\&D Program of China No. 2022YFF0504300;
National Natural Science Foundation of China (NSFC) grants 12361141814.

\bibliography{mybib}{}

\begin{thebibliography}{}
\expandafter\ifx\csname natexlab\endcsname\relax\def\natexlab#1{#1}\fi
\providecommand{\url}[1]{\href{#1}{#1}}
\providecommand{\dodoi}[1]{doi:~\href{http://doi.org/#1}{\nolinkurl{#1}}}
\providecommand{\doeprint}[1]{\href{http://ascl.net/#1}{\nolinkurl{http://ascl.net/#1}}}
\providecommand{\doarXiv}[1]{\href{https://arxiv.org/abs/#1}{\nolinkurl{https://arxiv.org/abs/#1}}}

\bibitem[{{Baldwin}(1955)}]{Baldwin1955MNRAS}
{Baldwin}, J.~E. 1955, \mnras, 115, 684, \dodoi{10.1093/mnras/115.6.684}

\bibitem[{{Bennett} {et~al.}(2013){Bennett}, {Larson}, {Weiland}, {Jarosik}, {Hinshaw}, {Odegard}, {Smith}, {Hill}, {Gold}, {Halpern}, {Komatsu}, {Nolta}, {Page}, {Spergel}, {Wollack}, {Dunkley}, {Kogut}, {Limon}, {Meyer}, {Tucker}, \& {Wright}}]{Bennett2013ApJS}
{Bennett}, C.~L., {Larson}, D., {Weiland}, J.~L., {et~al.} 2013, \apjs, 208, 20, \dodoi{10.1088/0067-0049/208/2/20}

\bibitem[{{Berkhuijsen} {et~al.}(1971){Berkhuijsen}, {Haslam}, \& {Salter}}]{Berkhuijsen1971A&A}
{Berkhuijsen}, E.~M., {Haslam}, C.~G.~T., \& {Salter}, C.~J. 1971, \aap, 14, 252

\bibitem[{{Borka}(2007)}]{Borka2007MNRAS}
{Borka}, V. 2007, \mnras, 376, 634, \dodoi{10.1111/j.1365-2966.2007.11499.x}

\bibitem[{{Breitschwerdt} \& {de Avillez}(2006)}]{Breitschwerdt2006AA}
{Breitschwerdt}, D., \& {de Avillez}, M.~A. 2006, \aap, 452, L1, \dodoi{10.1051/0004-6361:20064989}

\bibitem[{{Burke} {et~al.}(2019){Burke}, {Graham}, \& {Wilkinson}}]{Burke_Book2019}
{Burke}, B.~F., {Graham}, \& {Wilkinson}, P.~N. 2019, {An introduction to radio astronomy, 4th edition} ({Cambridge University Press})

\bibitem[{{Burns} \& {Hallinan}(2020)}]{Burns2020AAS}
{Burns}, J.~O., \& {Hallinan}, G. 2020, in American Astronomical Society Meeting Abstracts, Vol. 235, American Astronomical Society Meeting Abstracts \#235, 130.01

\bibitem[{{Centurion} \& {Vladilo}(1991)}]{Centurion1991ApJ}
{Centurion}, M., \& {Vladilo}, G. 1991, \apj, 372, 494, \dodoi{10.1086/169995}

\bibitem[{{Chen} {et~al.}(2021){Chen}, {Yan}, {Deng}, {Wu}, {Wu}, {Xu}, \& {Zhou}}]{Chen2021RSPTA}
{Chen}, X., {Yan}, J., {Deng}, L., {et~al.} 2021, Philosophical Transactions of the Royal Society of London Series A, 379, 20190566, \dodoi{10.1098/rsta.2019.0566}

\bibitem[{{Chen} {et~al.}(2024){Chen}, {Gao}, {Wu}, {Zhang}, {Wang}, {Liu}, {Zou}, {Deng}, {Gong}, {He}, {Li}, {Sun}, {Suo}, {Wang}, {Wu}, {Xu}, {Xu}, {Yue}, {Zhang}, {Zhou}, {Zhou}, {Zhu}, \& {Zhu}}]{Chen2024arXiv}
{Chen}, X., {Gao}, F., {Wu}, F., {et~al.} 2024, arXiv e-prints, arXiv:2403.16409, \dodoi{10.48550/arXiv.2403.16409}

\bibitem[{{Ciotti} \& {Bertin}(1999)}]{Ciotti_Bertin1999}
{Ciotti}, L., \& {Bertin}, G. 1999, \aap, 352, 447, \dodoi{10.48550/arXiv.astro-ph/9911078}

\bibitem[{{Condon} \& {Ransom}(2016)}]{Condon2016erabook}
{Condon}, J.~J., \& {Ransom}, S.~M. 2016, {Essential Radio Astronomy}

\bibitem[{{Cong} {et~al.}(2021){Cong}, {Yue}, {Xu}, {Huang}, {Zuo}, \& {Chen}}]{Cong2021APJ}
{Cong}, Y., {Yue}, B., {Xu}, Y., {et~al.} 2021, \apj, 914, 128, \dodoi{10.3847/1538-4357/abf55c}

\bibitem[{{Cong} {et~al.}(2022){Cong}, {Yue}, {Xu}, {Shi}, \& {Chen}}]{Cong2022ApJ}
{Cong}, Y., {Yue}, B., {Xu}, Y., {Shi}, Y., \& {Chen}, X. 2022, \apj, 940, 180, \dodoi{10.3847/1538-4357/ac9df7}

\bibitem[{{Cordes} \& {Lazio}(2002)}]{ne2001_I}
{Cordes}, J.~M., \& {Lazio}, T.~J.~W. 2002, arXiv e-prints, astro.
\newblock \doarXiv{astro-ph/0207156}

\bibitem[{{Cordes} \& {Lazio}(2003)}]{ne2001_II}
---. 2003, arXiv e-prints, astro.
\newblock \doarXiv{astro-ph/0301598}

\bibitem[{{Cox} \& {Reynolds}(1987)}]{Cox1987ARA&A}
{Cox}, D.~P., \& {Reynolds}, R.~J. 1987, \araa, 25, 303, \dodoi{10.1146/annurev.aa.25.090187.001511}

\bibitem[{{de Geus}(1992)}]{deGeus1992AA}
{de Geus}, E.~J. 1992, \aap, 262, 258

\bibitem[{{Dickinson}(2018)}]{Dickinson2018Galax}
{Dickinson}, C. 2018, Galaxies, 6, 56, \dodoi{10.3390/galaxies6020056}

\bibitem[{{Dickinson} {et~al.}(2003){Dickinson}, {Davies}, \& {Davis}}]{Dickinson2003MNRAS}
{Dickinson}, C., {Davies}, R.~D., \& {Davis}, R.~J. 2003, \mnras, 341, 369, \dodoi{10.1046/j.1365-8711.2003.06439.x}

\bibitem[{{Dowell} {et~al.}(2017){Dowell}, {Taylor}, {Schinzel}, {Kassim}, \& {Stovall}}]{Dowell2017MNRAS}
{Dowell}, J., {Taylor}, G.~B., {Schinzel}, F.~K., {Kassim}, N.~E., \& {Stovall}, K. 2017, \mnras, 469, 4537, \dodoi{10.1093/mnras/stx1136}

\bibitem[{{Eastwood} {et~al.}(2018){Eastwood}, {Anderson}, {Monroe}, {Hallinan}, {Barsdell}, {Bourke}, {Clark}, {Ellingson}, {Dowell}, {Garsden}, {Greenhill}, {Hartman}, {Kocz}, {Lazio}, {Price}, {Schinzel}, {Taylor}, {Vedantham}, {Wang}, \& {Woody}}]{Eastwood2018AJ}
{Eastwood}, M.~W., {Anderson}, M.~M., {Monroe}, R.~M., {et~al.} 2018, \aj, 156, 32, \dodoi{10.3847/1538-3881/aac721}

\bibitem[{{Egger} \& {Aschenbach}(1995)}]{Egger1995A&A}
{Egger}, R.~J., \& {Aschenbach}, B. 1995, \aap, 294, L25, \dodoi{10.48550/arXiv.astro-ph/9412086}

\bibitem[{{Foreman-Mackey} {et~al.}(2013){Foreman-Mackey}, {Hogg}, {Lang}, \& {Goodman}}]{emceee2013}
{Foreman-Mackey}, D., {Hogg}, D.~W., {Lang}, D., \& {Goodman}, J. 2013, \pasp, 125, 306, \dodoi{10.1086/670067}

\bibitem[{{Guzm{\'a}n} {et~al.}(2011){Guzm{\'a}n}, {May}, {Alvarez}, \& {Maeda}}]{Guzman2011A&A}
{Guzm{\'a}n}, A.~E., {May}, J., {Alvarez}, H., \& {Maeda}, K. 2011, \aap, 525, A138, \dodoi{10.1051/0004-6361/200913628}

\bibitem[{{Haffner} {et~al.}(2003){Haffner}, {Reynolds}, {Tufte}, {Madsen}, {Jaehnig}, \& {Percival}}]{Haffner2003ApJS}
{Haffner}, L.~M., {Reynolds}, R.~J., {Tufte}, S.~L., {et~al.} 2003, \apjs, 149, 405, \dodoi{10.1086/378850}

\bibitem[{{Hanbury Brown} {et~al.}(1960){Hanbury Brown}, {Davies}, \& {Hazard}}]{HanburyBrown1960Obs}
{Hanbury Brown}, R., {Davies}, R.~D., \& {Hazard}, C. 1960, The Observatory, 80, 191

\bibitem[{{Haslam} {et~al.}(1982){Haslam}, {Salter}, {Stoffel}, \& {Wilson}}]{Haslam1982A&A}
{Haslam}, C.~G.~T., {Salter}, C.~J., {Stoffel}, H., \& {Wilson}, W.~E. 1982, \aaps, 47, 1

\bibitem[{{Huang} {et~al.}(2019){Huang}, {Wu}, \& {Chen}}]{Huang2019SCPMA}
{Huang}, Q., {Wu}, F., \& {Chen}, X. 2019, Science China Physics, Mechanics, and Astronomy, 62, 989511, \dodoi{10.1007/s11433-018-9333-1}

\bibitem[{{Intema} {et~al.}(2017){Intema}, {Jagannathan}, {Mooley}, \& {Frail}}]{IntemaHT2017A&A}
{Intema}, H.~T., {Jagannathan}, P., {Mooley}, K.~P., \& {Frail}, D.~A. 2017, \aap, 598, A78, \dodoi{10.1051/0004-6361/201628536}

\bibitem[{{Irfan} {et~al.}(2022){Irfan}, {Bull}, {Santos}, {Wang}, {Grainge}, {Li}, {Carucci}, {Spinelli}, \& {Cunnington}}]{Irfan2022MNRAS}
{Irfan}, M.~O., {Bull}, P., {Santos}, M.~G., {et~al.} 2022, \mnras, 509, 4923, \dodoi{10.1093/mnras/stab3346}

\bibitem[{{Iwan}(1980)}]{Iwan1980ApJ}
{Iwan}, D. 1980, \apj, 239, 316, \dodoi{10.1086/158113}

\bibitem[{{Iwashita} {et~al.}(2023){Iwashita}, {Kataoka}, \& {Sofue}}]{Iwashita2023ApJ}
{Iwashita}, R., {Kataoka}, J., \& {Sofue}, Y. 2023, \apj, 958, 83, \dodoi{10.3847/1538-4357/ad0374}

\bibitem[{{Jansson} \& {Farrar}(2012)}]{Jansson2012ApJ}
{Jansson}, R., \& {Farrar}, G.~R. 2012, \apj, 757, 14, \dodoi{10.1088/0004-637X/757/1/14}

\bibitem[{{Kaaret} {et~al.}(2019){Kaaret}, {Zajczyk}, {LaRocca}, {Ringuette}, {Bluem}, {Fuelberth}, {Gulick}, {Jahoda}, {Johnson}, {Kirchner}, {Koutroumpa}, {Kuntz}, {McCurdy}, {Miles}, {Robison}, \& {Silich}}]{Kaaret2019ApJ}
{Kaaret}, P., {Zajczyk}, A., {LaRocca}, D.~M., {et~al.} 2019, \apj, 884, 162, \dodoi{10.3847/1538-4357/ab4193}

\bibitem[{{Kataoka} {et~al.}(2018){Kataoka}, {Sofue}, {Inoue}, {Akita}, {Nakashima}, \& {Totani}}]{Kataoka2018Galax}
{Kataoka}, J., {Sofue}, Y., {Inoue}, Y., {et~al.} 2018, Galaxies, 6, 27, \dodoi{10.3390/galaxies6010027}

\bibitem[{{Lallement}(2023)}]{Lallement2023CRPhy}
{Lallement}, R. 2023, Comptes Rendus Physique, 23, 1, \dodoi{10.5802/crphys.97}

\bibitem[{{Lallement} {et~al.}(2018){Lallement}, {Capitanio}, {Ruiz-Dern}, {Danielski}, {Babusiaux}, {Vergely}, {Elyajouri}, {Arenou}, \& {Leclerc}}]{Lallement2018A&A}
{Lallement}, R., {Capitanio}, L., {Ruiz-Dern}, L., {et~al.} 2018, \aap, 616, A132, \dodoi{10.1051/0004-6361/201832832}

\bibitem[{{Landecker} \& {Wielebinski}(1970)}]{Landecker1970AuJPA}
{Landecker}, T.~L., \& {Wielebinski}, R. 1970, Australian Journal of Physics Astrophysical Supplement, 16, 1

\bibitem[{{Large} {et~al.}(1966){Large}, {Quigley}, \& {Haslam}}]{Large1966MNRAS}
{Large}, M.~I., {Quigley}, M.~F.~S., \& {Haslam}, C.~G.~T. 1966, \mnras, 131, 335, \dodoi{10.1093/mnras/131.3.335}

\bibitem[{{Large} {et~al.}(1962){Large}, {Quigley}, \& {Haslam}}]{Large1962MNRAS}
{Large}, M.~I., {Quigley}, M.~J.~S., \& {Haslam}, C.~G.~T. 1962, \mnras, 124, 405, \dodoi{10.1093/mnras/124.5.405}

\bibitem[{{LaRocca} {et~al.}(2020{\natexlab{a}}){LaRocca}, {Kaaret}, {Kuntz}, {Hodges-Kluck}, {Zajczyk}, {Bluem}, {Ringuette}, \& {Jahoda}}]{LaRocca2020ApJ}
{LaRocca}, D.~M., {Kaaret}, P., {Kuntz}, K.~D., {et~al.} 2020{\natexlab{a}}, \apj, 904, 54, \dodoi{10.3847/1538-4357/abbdfd}

\bibitem[{{LaRocca} {et~al.}(2020{\natexlab{b}}){LaRocca}, {Kaaret}, {Kirchner}, {Zajczyk}, {Robison}, {Johnson}, {Jahoda}, {Fuelberth}, {Gulick}, {McCurdy}, {White}, \& {Miles}}]{LaRocca2020JATIS}
{LaRocca}, D.~M., {Kaaret}, P., {Kirchner}, D.~L., {et~al.} 2020{\natexlab{b}}, Journal of Astronomical Telescopes, Instruments, and Systems, 6, 014003, \dodoi{10.1117/1.JATIS.6.1.014003}

\bibitem[{{Liu} {et~al.}(2024){Liu}, {Merloni}, {Sanders}, {Ponti}, {Strong}, {Yeung}, {Locatelli}, {Predehl}, {Zheng}, {Sasaki}, {Freyberg}, {Dennerl}, {Becker}, {Nandra}, {Mayer}, \& {Buchner}}]{Liu2024ApJ}
{Liu}, T., {Merloni}, A., {Sanders}, J., {et~al.} 2024, \apjl, 967, L27, \dodoi{10.3847/2041-8213/ad47e0}

\bibitem[{{MacArthur} {et~al.}(2003){MacArthur}, {Courteau}, \& {Holtzman}}]{MacArthur2003ApJ}
{MacArthur}, L.~A., {Courteau}, S., \& {Holtzman}, J.~A. 2003, \apj, 582, 689, \dodoi{10.1086/344506}

\bibitem[{{Mertsch} \& {Sarkar}(2013)}]{Mertsch2013JCAP}
{Mertsch}, P., \& {Sarkar}, S. 2013, \jcap, 2013, 041, \dodoi{10.1088/1475-7516/2013/06/041}

\bibitem[{{Mou} {et~al.}(2023{\natexlab{a}}){Mou}, {Wu}, \& {Sofue}}]{Mou2023A&A}
{Mou}, G., {Wu}, J., \& {Sofue}, Y. 2023{\natexlab{a}}, \aap, 676, L3, \dodoi{10.1051/0004-6361/202245401}

\bibitem[{{Mou} {et~al.}(2023{\natexlab{b}}){Mou}, {Sun}, {Fang}, {Wang}, {Zhang}, {Yuan}, {Sofue}, {Wang}, \& {He}}]{Mou2023NatCo}
{Mou}, G., {Sun}, D., {Fang}, T., {et~al.} 2023{\natexlab{b}}, Nature Communications, 14, 781, \dodoi{10.1038/s41467-023-36478-0}

\bibitem[{{Padovani} {et~al.}(2021){Padovani}, {Bracco}, {Jeli{\'c}}, {Galli}, \& {Bellomi}}]{Padovani2021A&A}
{Padovani}, M., {Bracco}, A., {Jeli{\'c}}, V., {Galli}, D., \& {Bellomi}, E. 2021, \aap, 651, A116, \dodoi{10.1051/0004-6361/202140799}

\bibitem[{{Panopoulou} {et~al.}(2021){Panopoulou}, {Dickinson}, {Readhead}, {Pearson}, \& {Peel}}]{Panopoulou2021ApJ}
{Panopoulou}, G.~V., {Dickinson}, C., {Readhead}, A.~C.~S., {Pearson}, T.~J., \& {Peel}, M.~W. 2021, \apj, 922, 210, \dodoi{10.3847/1538-4357/ac273f}

\bibitem[{{Patra} {et~al.}(2015){Patra}, {Subrahmanyan}, {Sethi}, {Udaya Shankar}, \& {Raghunathan}}]{Patra2015ApJ}
{Patra}, N., {Subrahmanyan}, R., {Sethi}, S., {Udaya Shankar}, N., \& {Raghunathan}, A. 2015, \apj, 801, 138, \dodoi{10.1088/0004-637X/801/2/138}

\bibitem[{{Planck Collaboration} {et~al.}(2016{\natexlab{a}}){Planck Collaboration}, {Ade}, {Aghanim}, {Alves}, {Arnaud}, {Ashdown}, {Aumont}, {Baccigalupi}, {Banday}, {Barreiro}, {Bartlett}, {Bartolo}, {Battaner}, {Benabed}, {Beno{\^\i}t}, {Benoit-L{\'e}vy}, {Bernard}, {Bersanelli}, {Bielewicz}, {Bock}, {Bonaldi}, {Bonavera}, {Bond}, {Borrill}, {Bouchet}, {Boulanger}, {Bucher}, {Burigana}, {Butler}, {Calabrese}, {Cardoso}, {Catalano}, {Challinor}, {Chamballu}, {Chary}, {Chiang}, {Christensen}, {Colombi}, {Colombo}, {Combet}, {Couchot}, {Coulais}, {Crill}, {Curto}, {Cuttaia}, {Danese}, {Davies}, {Davis}, {de Bernardis}, {de Rosa}, {de Zotti}, {Delabrouille}, {Delouis}, {D{\'e}sert}, {Dickinson}, {Diego}, {Dole}, {Donzelli}, {Dor{\'e}}, {Douspis}, {Ducout}, {Dupac}, {Efstathiou}, {Elsner}, {En{\ss}lin}, {Eriksen}, {Falgarone}, {Fergusson}, {Finelli}, {Forni}, {Frailis}, {Fraisse}, {Franceschi}, {Frejsel}, {Galeotta}, {Galli}, {Ganga}, {Ghosh}, {Giard}, {Giraud-H{\'e}raud}, {Gjerl{\o}w}, {Gonz{\'a}lez-Nuevo},
  {G{\'o}rski}, {Gratton}, {Gregorio}, {Gruppuso}, {Gudmundsson}, {Hansen}, {Hanson}, {Harrison}, {Helou}, {Henrot-Versill{\'e}}, {Hern{\'a}ndez-Monteagudo}, {Herranz}, {Hildebrandt}, {Hivon}, {Hobson}, {Holmes}, {Hornstrup}, {Hovest}, {Huffenberger}, {Hurier}, {Jaffe}, {Jaffe}, {Jones}, {Juvela}, {Keih{\"a}nen}, {Keskitalo}, {Kisner}, {Kneissl}, {Knoche}, {Kunz}, {Kurki-Suonio}, {Lagache}, {L{\"a}hteenm{\"a}ki}, {Lamarre}, {Lasenby}, {Lattanzi}, {Lawrence}, {Leahy}, {Leonardi}, {Lesgourgues}, {Levrier}, {Liguori}, {Lilje}, {Linden-V{\o}rnle}, {L{\'o}pez-Caniego}, {Lubin}, {Mac{\'\i}as-P{\'e}rez}, {Maggio}, {Maino}, {Mandolesi}, {Mangilli}, {Maris}, {Marshall}, {Martin}, {Mart{\'\i}nez-Gonz{\'a}lez}, {Masi}, {Matarrese}, {McGehee}, {Meinhold}, {Melchiorri}, {Mendes}, {Mennella}, {Migliaccio}, {Mitra}, {Miville-Desch{\^e}nes}, {Moneti}, {Montier}, {Morgante}, {Mortlock}, {Moss}, {Munshi}, {Murphy}, {Nati}, {Natoli}, {Netterfield}, {N{\o}rgaard-Nielsen}, {Noviello}, {Novikov}, {Novikov}, {Orlando}, {Oxborrow},
  {Paci}, {Pagano}, {Pajot}, {Paladini}, {Paoletti}, {Partridge}, {Pasian}, {Patanchon}, {Pearson}, {Peel}, {Perdereau}, {Perotto}, {Perrotta}, {Pettorino}, {Piacentini}, {Piat}, {Pierpaoli}, {Pietrobon}, {Plaszczynski}, {Pointecouteau}, {Polenta}, {Pratt}, {Pr{\'e}zeau}, {Prunet}, {Puget}, {Rachen}, {Reach}, {Rebolo}, {Reinecke}, {Remazeilles}, {Renault}, {Renzi}, {Ristorcelli}, {Rocha}, {Rosset}, {Rossetti}, {Roudier}, {Rubi{\~n}o-Mart{\'\i}n}, {Rusholme}, {Sandri}, {Santos}, {Savelainen}, {Savini}, {Scott}, {Seiffert}, {Shellard}, {Spencer}, {Stolyarov}, {Stompor}, {Strong}, {Sudiwala}, {Sunyaev}, {Sutton}, {Suur-Uski}, {Sygnet}, {Tauber}, {Terenzi}, {Toffolatti}, {Tomasi}, {Tristram}, {Tucci}, {Tuovinen}, {Umana}, {Valenziano}, {Valiviita}, {Van Tent}, {Vidal}, {Vielva}, {Villa}, {Wade}, {Wandelt}, {Watson}, {Wehus}, {Wilkinson}, {Yvon}, {Zacchei}, \& {Zonca}}]{Planck2016XXV}
{Planck Collaboration}, {Ade}, P.~A.~R., {Aghanim}, N., {et~al.} 2016{\natexlab{a}}, \aap, 594, A25, \dodoi{10.1051/0004-6361/201526803}

\bibitem[{{Planck Collaboration} {et~al.}(2016{\natexlab{b}}){Planck Collaboration}, {Adam}, {Ade}, {Aghanim}, {Alves}, {Arnaud}, {Ashdown}, {Aumont}, {Baccigalupi}, {Banday}, {Barreiro}, {Bartlett}, {Bartolo}, {Battaner}, {Benabed}, {Beno{\^\i}t}, {Benoit-L{\'e}vy}, {Bernard}, {Bersanelli}, {Bielewicz}, {Bock}, {Bonaldi}, {Bonavera}, {Bond}, {Borrill}, {Bouchet}, {Boulanger}, {Bucher}, {Burigana}, {Butler}, {Calabrese}, {Cardoso}, {Catalano}, {Challinor}, {Chamballu}, {Chary}, {Chiang}, {Christensen}, {Clements}, {Colombi}, {Colombo}, {Combet}, {Couchot}, {Coulais}, {Crill}, {Curto}, {Cuttaia}, {Danese}, {Davies}, {Davis}, {de Bernardis}, {de Rosa}, {de Zotti}, {Delabrouille}, {D{\'e}sert}, {Dickinson}, {Diego}, {Dole}, {Donzelli}, {Dor{\'e}}, {Douspis}, {Ducout}, {Dupac}, {Efstathiou}, {Elsner}, {En{\ss}lin}, {Eriksen}, {Falgarone}, {Fergusson}, {Finelli}, {Forni}, {Frailis}, {Fraisse}, {Franceschi}, {Frejsel}, {Galeotta}, {Galli}, {Ganga}, {Ghosh}, {Giard}, {Giraud-H{\'e}raud}, {Gjerl{\o}w},
  {Gonz{\'a}lez-Nuevo}, {G{\'o}rski}, {Gratton}, {Gregorio}, {Gruppuso}, {Gudmundsson}, {Hansen}, {Hanson}, {Harrison}, {Helou}, {Henrot-Versill{\'e}}, {Hern{\'a}ndez-Monteagudo}, {Herranz}, {Hildebrandt}, {Hivon}, {Hobson}, {Holmes}, {Hornstrup}, {Hovest}, {Huffenberger}, {Hurier}, {Jaffe}, {Jaffe}, {Jones}, {Juvela}, {Keih{\"a}nen}, {Keskitalo}, {Kisner}, {Kneissl}, {Knoche}, {Kunz}, {Kurki-Suonio}, {Lagache}, {L{\"a}hteenm{\"a}ki}, {Lamarre}, {Lasenby}, {Lattanzi}, {Lawrence}, {Le Jeune}, {Leahy}, {Leonardi}, {Lesgourgues}, {Levrier}, {Liguori}, {Lilje}, {Linden-V{\o}rnle}, {L{\'o}pez-Caniego}, {Lubin}, {Mac{\'\i}as-P{\'e}rez}, {Maggio}, {Maino}, {Mandolesi}, {Mangilli}, {Maris}, {Marshall}, {Martin}, {Mart{\'\i}nez-Gonz{\'a}lez}, {Masi}, {Matarrese}, {McGehee}, {Meinhold}, {Melchiorri}, {Mendes}, {Mennella}, {Migliaccio}, {Mitra}, {Miville-Desch{\^e}nes}, {Moneti}, {Montier}, {Morgante}, {Mortlock}, {Moss}, {Munshi}, {Murphy}, {Naselsky}, {Nati}, {Natoli}, {Netterfield}, {N{\o}rgaard-Nielsen}, {Noviello},
  {Novikov}, {Novikov}, {Orlando}, {Oxborrow}, {Paci}, {Pagano}, {Pajot}, {Paladini}, {Paoletti}, {Partridge}, {Pasian}, {Patanchon}, {Pearson}, {Perdereau}, {Perotto}, {Perrotta}, {Pettorino}, {Piacentini}, {Piat}, {Pierpaoli}, {Pietrobon}, {Plaszczynski}, {Pointecouteau}, {Polenta}, {Pratt}, {Pr{\'e}zeau}, {Prunet}, {Puget}, {Rachen}, {Reach}, {Rebolo}, {Reinecke}, {Remazeilles}, {Renault}, {Renzi}, {Ristorcelli}, {Rocha}, \& {Rosset}}]{Plank2016A&A}
{Planck Collaboration}, {Adam}, R., {Ade}, P.~A.~R., {et~al.} 2016{\natexlab{b}}, \aap, 594, A10, \dodoi{10.1051/0004-6361/201525967}

\bibitem[{{Planck Collaboration} {et~al.}(2020){Planck Collaboration}, {Aghanim}, {Akrami}, {Arroja}, {Ashdown}, {Aumont}, {Baccigalupi}, {Ballardini}, {Banday}, {Barreiro}, {Bartolo}, {Basak}, {Battye}, {Benabed}, {Bernard}, {Bersanelli}, {Bielewicz}, {Bock}, {Bond}, {Borrill}, {Bouchet}, {Boulanger}, {Bucher}, {Burigana}, {Butler}, {Calabrese}, {Cardoso}, {Carron}, {Casaponsa}, {Challinor}, {Chiang}, {Colombo}, {Combet}, {Contreras}, {Crill}, {Cuttaia}, {de Bernardis}, {de Zotti}, {Delabrouille}, {Delouis}, {D{\'e}sert}, {Di Valentino}, {Dickinson}, {Diego}, {Donzelli}, {Dor{\'e}}, {Douspis}, {Ducout}, {Dupac}, {Efstathiou}, {Elsner}, {En{\ss}lin}, {Eriksen}, {Falgarone}, {Fantaye}, {Fergusson}, {Fernandez-Cobos}, {Finelli}, {Forastieri}, {Frailis}, {Franceschi}, {Frolov}, {Galeotta}, {Galli}, {Ganga}, {G{\'e}nova-Santos}, {Gerbino}, {Ghosh}, {Gonz{\'a}lez-Nuevo}, {G{\'o}rski}, {Gratton}, {Gruppuso}, {Gudmundsson}, {Hamann}, {Handley}, {Hansen}, {Helou}, {Herranz}, {Hildebrandt}, {Hivon}, {Huang}, {Jaffe},
  {Jones}, {Karakci}, {Keih{\"a}nen}, {Keskitalo}, {Kiiveri}, {Kim}, {Kisner}, {Knox}, {Krachmalnicoff}, {Kunz}, {Kurki-Suonio}, {Lagache}, {Lamarre}, {Langer}, {Lasenby}, {Lattanzi}, {Lawrence}, {Le Jeune}, {Leahy}, {Lesgourgues}, {Levrier}, {Lewis}, {Liguori}, {Lilje}, {Lilley}, {Lindholm}, {L{\'o}pez-Caniego}, {Lubin}, {Ma}, {Mac{\'\i}as-P{\'e}rez}, {Maggio}, {Maino}, {Mandolesi}, {Mangilli}, {Marcos-Caballero}, {Maris}, {Martin}, {Martinelli}, {Mart{\'\i}nez-Gonz{\'a}lez}, {Matarrese}, {Mauri}, {McEwen}, {Meerburg}, {Meinhold}, {Melchiorri}, {Mennella}, {Migliaccio}, {Millea}, {Mitra}, {Miville-Desch{\^e}nes}, {Molinari}, {Moneti}, {Montier}, {Morgante}, {Moss}, {Mottet}, {M{\"u}nchmeyer}, {Natoli}, {N{\o}rgaard-Nielsen}, {Oxborrow}, {Pagano}, {Paoletti}, {Partridge}, {Patanchon}, {Pearson}, {Peel}, {Peiris}, {Perrotta}, {Pettorino}, {Piacentini}, {Polastri}, {Polenta}, {Puget}, {Rachen}, {Reinecke}, {Remazeilles}, {Renault}, {Renzi}, {Rocha}, {Rosset}, {Roudier}, {Rubi{\~n}o-Mart{\'\i}n},
  {Ruiz-Granados}, {Salvati}, {Sandri}, {Savelainen}, {Scott}, {Shellard}, {Shiraishi}, {Sirignano}, {Sirri}, {Spencer}, {Sunyaev}, {Suur-Uski}, {Tauber}, {Tavagnacco}, {Tenti}, {Terenzi}, {Toffolatti}, {Tomasi}, {Trombetti}, {Valiviita}, {Van Tent}, {Vibert}, {Vielva}, {Villa}, {Vittorio}, {Wandelt}, {Wehus}, {White}, {White}, {Zacchei}, \& {Zonca}}]{Plank2020I}
{Planck Collaboration}, {Aghanim}, N., {Akrami}, Y., {et~al.} 2020, \aap, 641, A1, \dodoi{10.1051/0004-6361/201833880}

\bibitem[{{Predehl} {et~al.}(2020){Predehl}, {Sunyaev}, {Becker}, {Brunner}, {Burenin}, {Bykov}, {Cherepashchuk}, {Chugai}, {Churazov}, {Doroshenko}, {Eismont}, {Freyberg}, {Gilfanov}, {Haberl}, {Khabibullin}, {Krivonos}, {Maitra}, {Medvedev}, {Merloni}, {Nandra}, {Nazarov}, {Pavlinsky}, {Ponti}, {Sanders}, {Sasaki}, {Sazonov}, {Strong}, \& {Wilms}}]{Predehl2020Natur}
{Predehl}, P., {Sunyaev}, R.~A., {Becker}, W., {et~al.} 2020, \nat, 588, 227, \dodoi{10.1038/s41586-020-2979-0}

\bibitem[{Quigley \& Haslam(1965)}]{Quigley1965StructureOT}
Quigley, M. J.~S., \& Haslam, C. G.~T. 1965, Nature, 208, 741.
\newblock \url{https://api.semanticscholar.org/CorpusID:4205652}

\bibitem[{{Reich} \& {Reich}(1986)}]{Reich1986A&A}
{Reich}, P., \& {Reich}, W. 1986, \aaps, 63, 205

\bibitem[{{Reich} \& {Reich}(2009)}]{Reich2009IAUS}
{Reich}, W., \& {Reich}, P. 2009, in IAU Symposium, Vol. 259, Cosmic Magnetic Fields: From Planets, to Stars and Galaxies, ed. K.~G. {Strassmeier}, A.~G. {Kosovichev}, \& J.~E. {Beckman}, 603--612, \dodoi{10.1017/S1743921309031433}

\bibitem[{{Reis} \& {Corradi}(2008)}]{Reis2008AA}
{Reis}, W., \& {Corradi}, W.~J.~B. 2008, \aap, 486, 471, \dodoi{10.1051/0004-6361:20077946}

\bibitem[{{Remazeilles} {et~al.}(2015){Remazeilles}, {Dickinson}, {Banday}, {Bigot-Sazy}, \& {Ghosh}}]{Remazeilles2015MNRAS}
{Remazeilles}, M., {Dickinson}, C., {Banday}, A.~J., {Bigot-Sazy}, M.~A., \& {Ghosh}, T. 2015, \mnras, 451, 4311, \dodoi{10.1093/mnras/stv1274}

\bibitem[{{Roger} {et~al.}(1999){Roger}, {Costain}, {Landecker}, \& {Swerdlyk}}]{Roger1999A&A}
{Roger}, R.~S., {Costain}, C.~H., {Landecker}, T.~L., \& {Swerdlyk}, C.~M. 1999, \aaps, 137, 7, \dodoi{10.1051/aas:1999239}

\bibitem[{{Sallmen} {et~al.}(2008){Sallmen}, {Korpela}, \& {Yamashita}}]{Sallmen2008ApJ}
{Sallmen}, S.~M., {Korpela}, E.~J., \& {Yamashita}, H. 2008, \apj, 681, 1310, \dodoi{10.1086/588802}

\bibitem[{{Santos} {et~al.}(2011){Santos}, {Corradi}, \& {Reis}}]{Santos2011ApJ}
{Santos}, F.~P., {Corradi}, W., \& {Reis}, W. 2011, \apj, 728, 104, \dodoi{10.1088/0004-637X/728/2/104}

\bibitem[{{Sarkar}(2019)}]{Sarkar2019MNRAS}
{Sarkar}, K.~C. 2019, \mnras, 482, 4813, \dodoi{10.1093/mnras/sty2944}

\bibitem[{{Seiffert} {et~al.}(2011){Seiffert}, {Fixsen}, {Kogut}, {Levin}, {Limon}, {Lubin}, {Mirel}, {Singal}, {Villela}, {Wollack}, \& {Wuensche}}]{Seiffert2011ApJ}
{Seiffert}, M., {Fixsen}, D.~J., {Kogut}, A., {et~al.} 2011, \apj, 734, 6, \dodoi{10.1088/0004-637X/734/1/6}

\bibitem[{{Sersic}(1968)}]{sersic1968}
{Sersic}, J.~L. 1968, {Atlas de Galaxias Australes}

\bibitem[{{Shchekinov}(2018)}]{Shchekinov2018Galax}
{Shchekinov}, Y. 2018, Galaxies, 6, 62, \dodoi{10.3390/galaxies6020062}

\bibitem[{{Sofue}(1977)}]{Sofue1977A&A}
{Sofue}, Y. 1977, \aap, 60, 327

\bibitem[{{Sofue}(1994)}]{Sofue1994ApJ}
---. 1994, \apjl, 431, L91, \dodoi{10.1086/187480}

\bibitem[{{Sofue}(2000)}]{Sofue2000ApJ}
---. 2000, \apj, 540, 224, \dodoi{10.1086/309297}

\bibitem[{{Sofue}(2015)}]{Sofue2015MNRAS}
---. 2015, \mnras, 447, 3824, \dodoi{10.1093/mnras/stu2661}

\bibitem[{{Sofue}(2017)}]{Sofue2017PASJ}
---. 2017, \pasj, 69, L8, \dodoi{10.1093/pasj/psx067}

\bibitem[{{Sofue} {et~al.}(2016){Sofue}, {Habe}, {Kataoka}, {Totani}, {Inoue}, {Nakashima}, {Matsui}, \& {Akita}}]{Sofue2016MNRAS}
{Sofue}, Y., {Habe}, A., {Kataoka}, J., {et~al.} 2016, \mnras, 459, 108, \dodoi{10.1093/mnras/stw623}

\bibitem[{{Sofue} {et~al.}(1974){Sofue}, {Hamajima}, \& {Fujimoto}}]{Sofue1974PASJ}
{Sofue}, Y., {Hamajima}, K., \& {Fujimoto}, M. 1974, \pasj, 26, 399

\bibitem[{{Sofue} {et~al.}(2023){Sofue}, {Kataoka}, \& {Iwashita}}]{Sofue2023MNRAS}
{Sofue}, Y., {Kataoka}, J., \& {Iwashita}, R. 2023, \mnras, 524, 4212, \dodoi{10.1093/mnras/stad1985}

\bibitem[{{Su} {et~al.}(2010){Su}, {Slatyer}, \& {Finkbeiner}}]{SuMeng2010ApJ}
{Su}, M., {Slatyer}, T.~R., \& {Finkbeiner}, D.~P. 2010, \apj, 724, 1044, \dodoi{10.1088/0004-637X/724/2/1044}

\bibitem[{{Sun} {et~al.}(2008){Sun}, {Reich}, {Waelkens}, \& {En{\ss}lin}}]{Sun2008A&A}
{Sun}, X.~H., {Reich}, W., {Waelkens}, A., \& {En{\ss}lin}, T.~A. 2008, \aap, 477, 573, \dodoi{10.1051/0004-6361:20078671}

\bibitem[{{Sun} {et~al.}(2015){Sun}, {Landecker}, {Gaensler}, {Carretti}, {Reich}, {Leahy}, {McClure-Griffiths}, {Crocker}, {Wolleben}, {Haverkorn}, {Douglas}, \& {Gray}}]{Sun2015ApJ}
{Sun}, X.~H., {Landecker}, T.~L., {Gaensler}, B.~M., {et~al.} 2015, \apj, 811, 40, \dodoi{10.1088/0004-637X/811/1/40}

\bibitem[{{Vidal} {et~al.}(2015){Vidal}, {Dickinson}, {Davies}, \& {Leahy}}]{Vidal2015MNRAS}
{Vidal}, M., {Dickinson}, C., {Davies}, R.~D., \& {Leahy}, J.~P. 2015, \mnras, 452, 656, \dodoi{10.1093/mnras/stv1328}

\bibitem[{{Vieira} {et~al.}(2023){Vieira}, {Korchagin}, {Carraro}, \& {Lutsenko}}]{Vieira2023Galax}
{Vieira}, K., {Korchagin}, V., {Carraro}, G., \& {Lutsenko}, A. 2023, Galaxies, 11, 77, \dodoi{10.3390/galaxies11030077}

\bibitem[{{Wolleben}(2007)}]{Wolleben2007ApJ}
{Wolleben}, M. 2007, \apj, 664, 349, \dodoi{10.1086/518711}

\bibitem[{{Yao} {et~al.}(2017){Yao}, {Manchester}, \& {Wang}}]{Yao2017ApJ}
{Yao}, J.~M., {Manchester}, R.~N., \& {Wang}, N. 2017, \apj, 835, 29, \dodoi{10.3847/1538-4357/835/1/29}

\bibitem[{{Zhang} {et~al.}(2024){Zhang}, {Ponti}, {Carretti}, {Liu}, {Morris}, {Haverkorn}, {Locatelli}, {Zheng}, {Aharonian}, {Zhang}, {Zhang}, {Stel}, {Strong}, {Yeung}, \& {Merloni}}]{Zhang2024NatAs}
{Zhang}, H.-S., {Ponti}, G., {Carretti}, E., {et~al.} 2024, Nature Astronomy, 8, 1416, \dodoi{10.1038/s41550-024-02362-0}

\end{thebibliography}
\bibliographystyle{aasjournal}



%\appendix  
%\renewcommand{\thefigure}{A.\arabic{figure}} % 重新定义图形编号格式  
%\setcounter{figure}{0} % 重置图形计数器

\end{document}